\documentclass[ %
 reprint,
superscriptaddress,
 amsmath,amssymb,
 aps,
prb,
]{revtex4-1}

\usepackage{graphicx}% Include figure files
\usepackage{dcolumn}% Align table columns on decimal point
\usepackage{bm}% bold math
\usepackage{subfigure}
\usepackage{epstopdf}
\usepackage{lettrine}

\begin{document}

\preprint{APS/123-QED}

\title{Fano Resonance in a cavity-reflector hybrid system}% Force line breaks with \\

\author{Chengyu Yan}
 \email{uceeya3@ucl.ac.uk}
 \affiliation{%
 London Centre for Nanotechnology, 17-19 Gordon Street, London WC1H 0AH, United Kingdom\\
 }%
 \affiliation{
  Department of Electronic and Electrical Engineering, University College London, Torrington Place, London WC1E 7JE, United Kingdom
 }%
 \author{Sanjeev Kumar}
 \affiliation{%
 London Centre for Nanotechnology, 17-19 Gordon Street, London WC1H 0AH, United Kingdom\\
 }%
 \affiliation{
  Department of Electronic and Electrical Engineering, University College London, Torrington Place, London WC1E 7JE, United Kingdom
 }%
\author{Michael Pepper}
\affiliation{%
 London Centre for Nanotechnology, 17-19 Gordon Street, London WC1H 0AH, United Kingdom\\
 }%
 \affiliation{
  Department of Electronic and Electrical Engineering, University College London, Torrington Place, London WC1E 7JE, United Kingdom
 }%
\author{Patrick See}
\affiliation{%
 National Physical Laboratory, Hampton Road, Teddington, Middlesex TW11 0LW, United Kingdom\\
}%
\author{Ian Farrer}
\affiliation{%
	Cavendish Laboratory, J.J. Thomson Avenue, Cambridge CB3 OHE, United Kingdom\\
}%
\author{David Ritchie}
\affiliation{%
 Cavendish Laboratory, J.J. Thomson Avenue, Cambridge CB3 OHE, United Kingdom\\
}%
\author{Jonathan Griffiths}
\affiliation{%
 Cavendish Laboratory, J.J. Thomson Avenue, Cambridge CB3 OHE, United Kingdom\\
}%
\author{Geraint Jones}
\affiliation{%
 Cavendish Laboratory, J.J. Thomson Avenue, Cambridge CB3 OHE, United Kingdom\\
}%

\date{\today}% It is always \today, today,
             
\begin{abstract}
We present the results of transport measurements in a hybrid system consisting of an arch-shaped quantum point contact (QPC) and a reflector; together, they form an electronic cavity in between them. On tuning the arch-QPC and the reflector, asymmetric resonance peak in resistance is observed at the one-dimension to two-dimension transition. Moreover, a dip in resistance near the pinch-off of the QPC is found to be strongly dependent on the reflector voltage. These two structures fit very well with the Fano line shape. The Fano resonance was found to get weakened on applying a transverse magnetic field, and it smeared out at 100 mT. In addition, the Fano like shape exhibited a strong temperature dependence and gradually smeared out when the temperature was increased from 1.5 to 20 K.  The results might be useful in realising device for quantum information processing.        
\end{abstract}

\maketitle

There is a growing interest in realising the electrical analog of photonic-cavity devices in condensed matter as they form the fundamental basis of tomorrow's quantum technologies for realising building blocks for quantum information processing and quantum circuits. Much success with photonic cavity\cite{RNO99,HBW07,NDH10,DK04} based system could be attributed to a better control over manipulating photons and achieving entanglement over a larger distance for information processing\cite{KMW95,YSJ04,CSK13,ESR10}. Electronic devices have not enjoyed a similar level of success due to limited advancement in controlling the quantum states of electrons. Recently, an electronic cavity coupled to a quantum dot was demonstrated with the observation of spin coherent state in the regime of dot-cavity coupling due to Kondo effect\cite{CDO15}. There have been experimental attempts in the past in coupling an electronic cavity with one-dimensional (1D) electrons using a quantum point contact (QPC)\cite{DTM88,WHB88}, however, in such cases the oscillations in the conductance\cite{MTM94,KMW97,DTW01,DFP05} were primarily due to quantum interference rather than coupling between the 1D-cavity states. Therefore, direct evidence of coupling between the 1D states of a QPC and the cavity states would be crucial in creating a platform for realising cavity based tunable electronic devices. 

In the present work, we demonstrate a cavity-reflector hybrid quantum device consisting of a QPC as the source of 1D electrons and an electronic cavity of 2D electrons, and show that when these states couple, they give rise to a Fano resonance\cite{YSM161}.

\begin{figure}
	\subfigure{    
		\includegraphics[height=1.8in,width=3.2in]{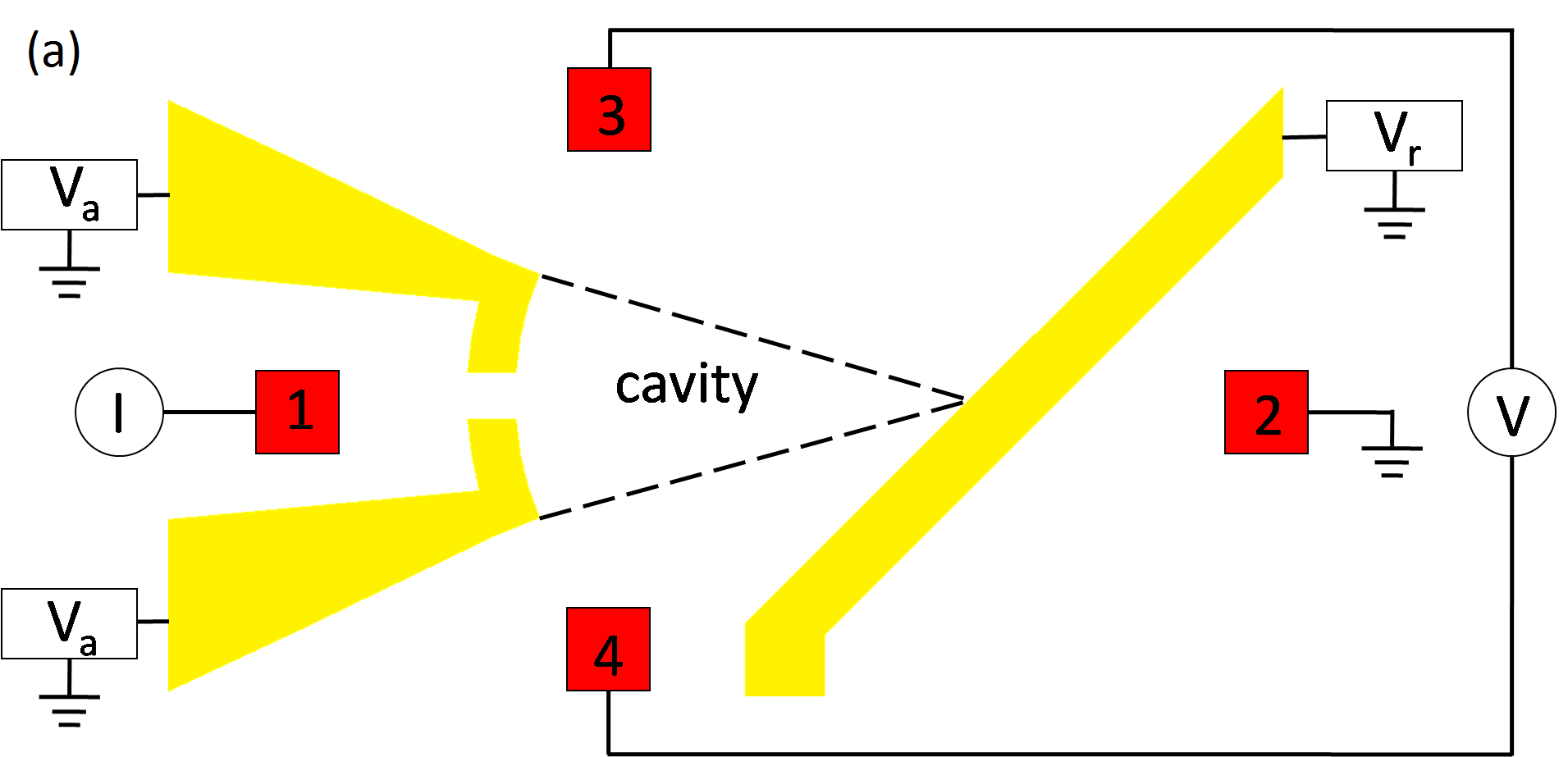} 
		\label{fig:Fig1A} 
	} % 
	             
	\subfigure{
		\includegraphics[height=2.0in,width=3.6in]{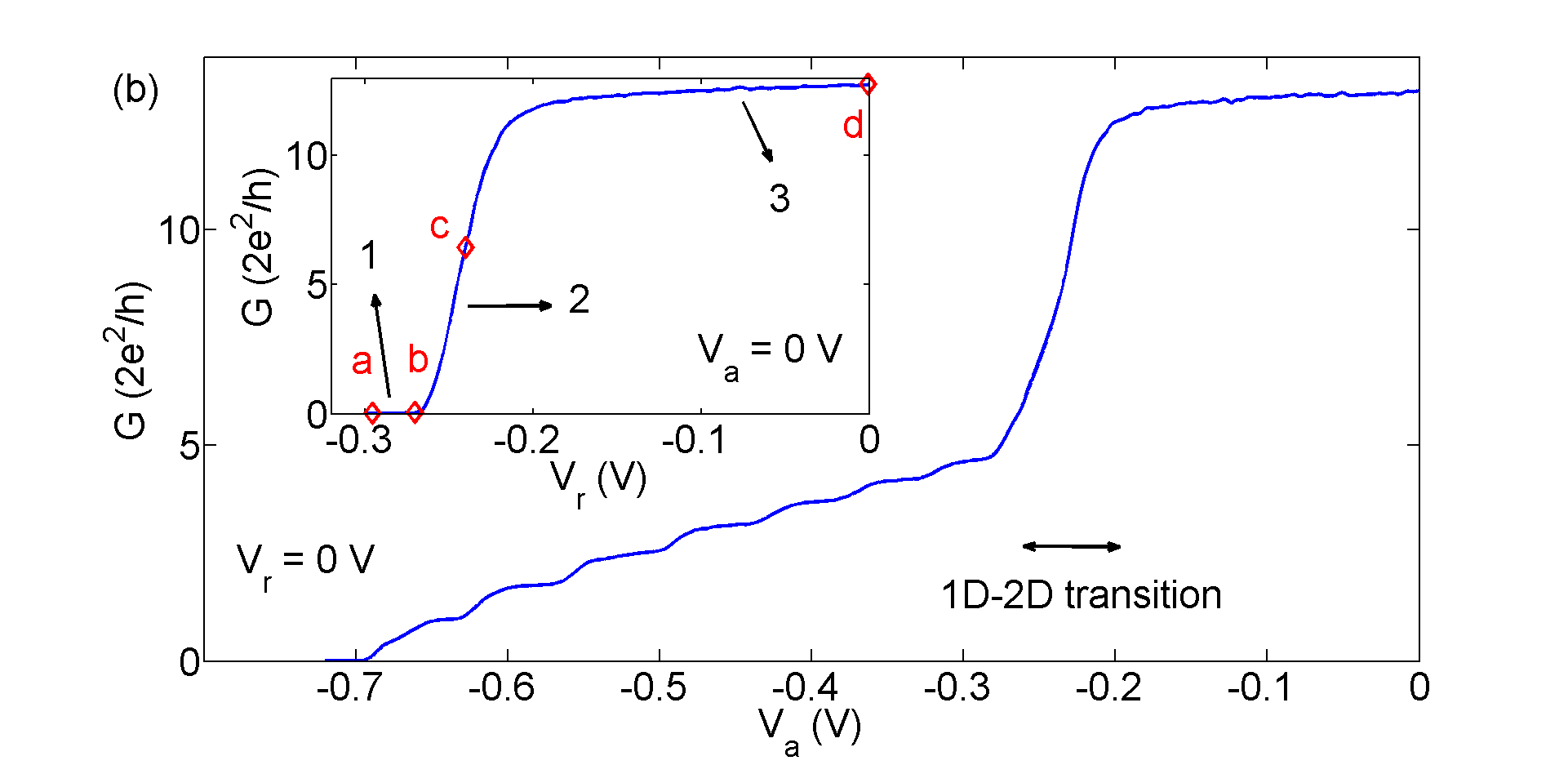}
		\label{fig:Fig1B}
	}% 
	
	\caption{Schematic of the device and the experiment setup. (a) the yellow blocks are metallic gates and red squares are Ohmics; excitation current is fed to Ohmic 1 while 2 is grounded; Ohmics 3 and 4 are voltage probes. The opening angle of the arch is 45$^\circ$ and the radius (also the distance between arch-gate and reflector) is 1.5 $\mu$m, both the length and width of the QPC embedded in the arch is 200 nm,  the length of inclined surface of the reflector is 3.0 $\mu$m and the width is 300 nm. (b) Differential conductance measurement of the QPC (main plot) and the reflector (inset). To be noted that in this plot when the QPC was measured, the reflector was grounded, and visa-versa. The voltage applied to the QPC (reflector) was V$_a$ (V$_r$). Three regimes are identified from the reflector voltage characteristic and labelled as regime 1 (a-b), regime 2 (b-c) and regime 3 (c-d). }           
	\label{fig:exp_setup}
\end{figure}  

 \begin{figure}
 	
 	\subfigure{    
 		\includegraphics[height=2.4in,width=3.6in]{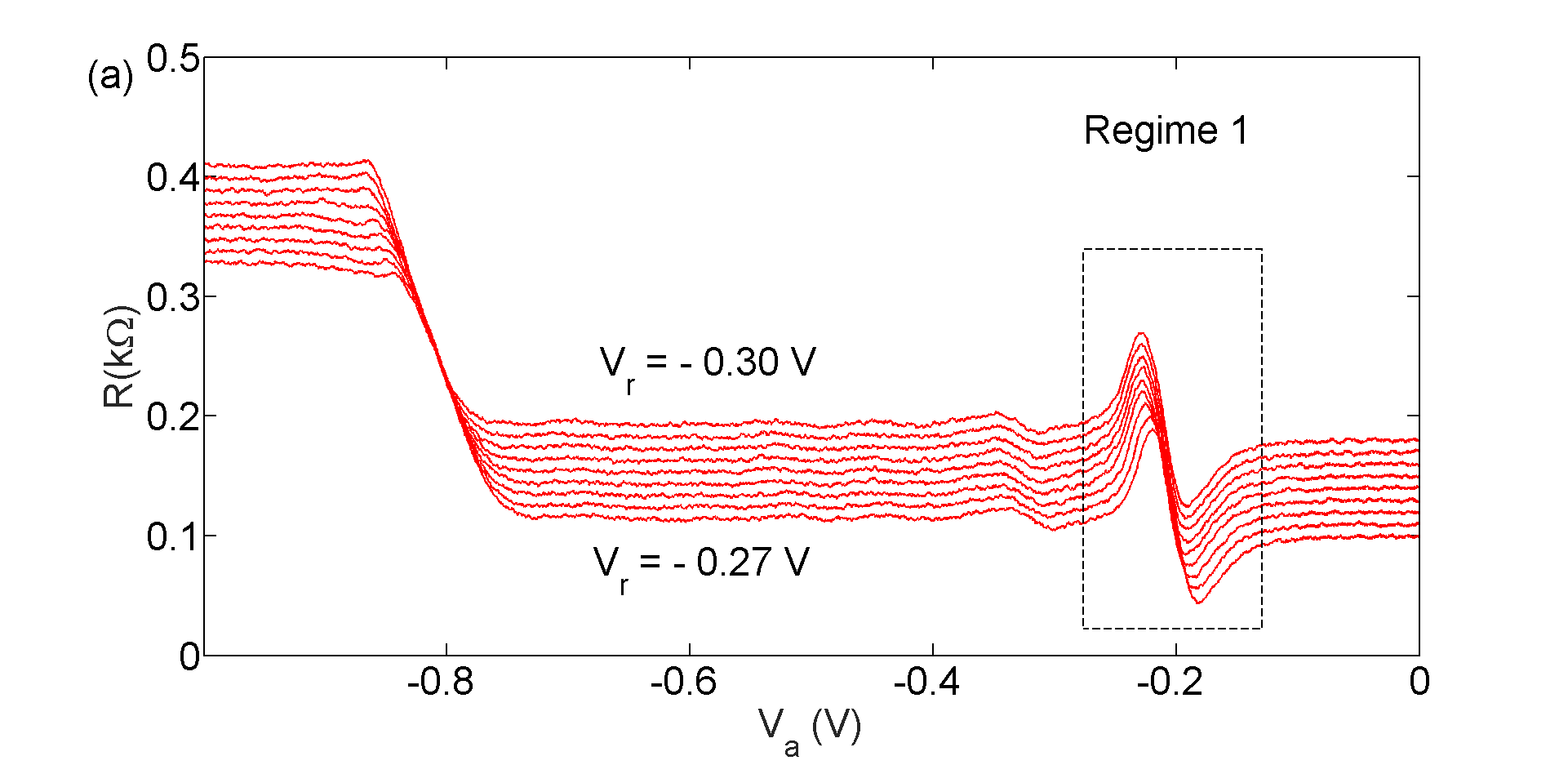} 
 		\label{fig:Fig2A } 
 	} % 
 	
 	\subfigure{
 		\includegraphics[height=2.4in,width=3.6in]{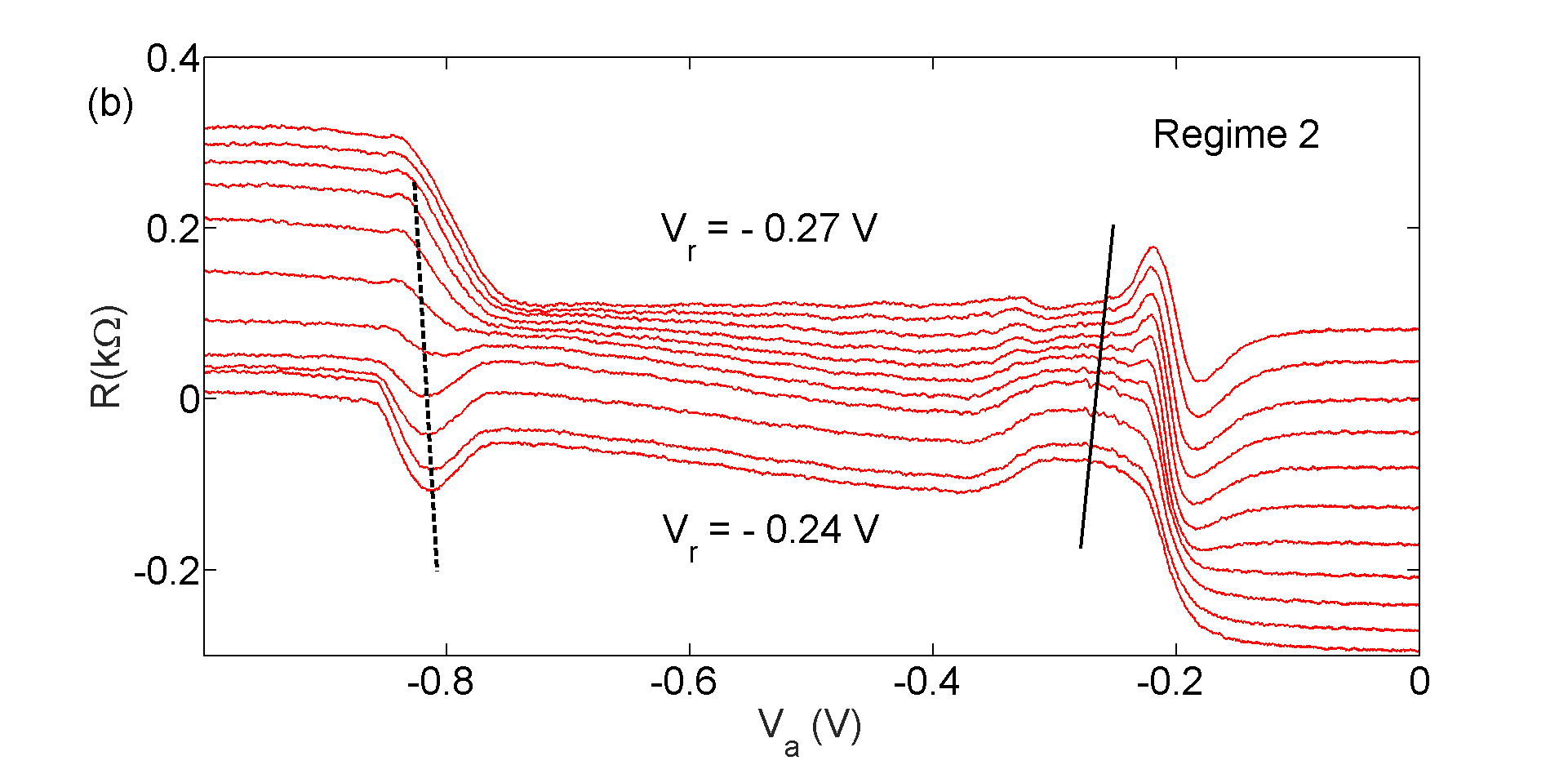}
 		\label{fig:Fig2B}
 	}% 
 	
 	\subfigure{
 		\includegraphics[height=2.4in,width=3.6in]{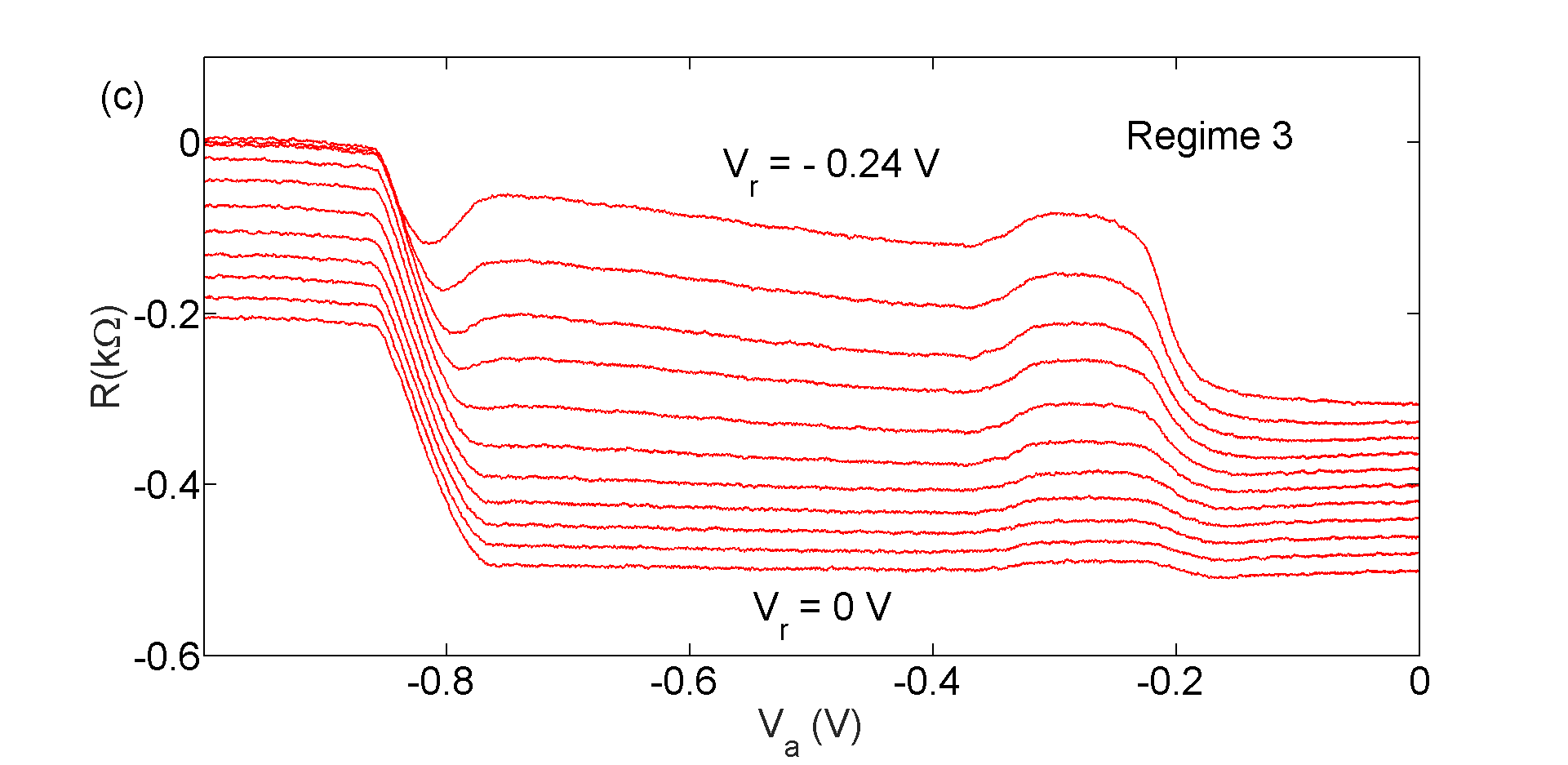}
 		\label{fig:Fig2C}
 	}%
 	
 	\caption[\textit{R} as a function of arch-QPC voltage for various V$_r$ ]{\textit{R} as a function of arch-QPC voltage for various V$_r$. (a) Result in regime 1: a strong asymmetric resonance structure highlighted by a dotted rectangle  occurs around - 0.2 V. (b) Result in regime 2: the resonance structure slowly evolves into a broad shoulder structure (marked by the solid line), meanwhile a dip gradually forms at the pinch-off regime of the arch-QPC (marked by the dashed line). (c) Result in regime 3: intensity of both the shoulder structure and the dip decreases with less negative reflector voltage. Data have been offset vertically by 10 $\Omega$ in each plots for clarity.   } 
 	\label{fig:cav_highT}
 \end{figure}

 The samples studied in the present work consist of a pair of arch-shaped gates, with QPC forming in the centre of the arch, and a reflector inclined at 75$^\circ$ against the current flow direction such that centre of the QPC and the reflector are aligned  as shown in Fig.~\ref{fig:exp_setup}(a)\cite{YSM162}. With this geometry the interference between incident and reflected electrons which contribute to oscillations reported previously\cite{KMW97,HHH99,HHH00} is significantly reduced because the incident and reflected electrons are spatially separated, thus all the features observed here are due to the coupling effect. The hybrid devices were fabricated on a high mobility two-dimensional electron gas (2DEG) formed at the interface of GaAs/Al$_{0.33}$Ga$_{0.67}$As heterostructure. The 2DEG is situated 90 nm from the surface where the gates are deposited. The electron density (mobility) measured at 1.5 K was 1.80$\times$10$^{11}$cm$^{-2}$ (2.1$\times$10$^6$cm$^2$V$^{-1}$s$^{-1}$) therefore both the mean free path and phase coherence length were over 10 $\mu$m which is much larger than the distance between the QPC and the reflector (1.5 $\mu$m). All the measurements were performed with standard lock-in technique in a cryofree pulsed-tube cryostat with a base temperature of 1.5 K. For the four terminal resistance measurement, a 10 nA at 77 Hz ac current was applied while an ac voltage of 10 $\mu$V at 77 Hz was used for two terminal conductance measurement. Figure~\ref{fig:exp_setup}(b) shows the conductance plot of the QPC with well defined conductance plateaus; on the other hand, the conductance of the reflector drops, when measured separately, around - 0.2 V, which indicates a sharp change in the transmission probability (Fig.~\ref{fig:exp_setup}(b), inset). The three regimes identified from the characteristic of the reflector are labelled as regime 1-3, as shown in the inset.

To highlight the effect of the cavity states, we studied the non-local four terminal resistance $R_{12,34} =  V_{34} / I_{12}$ (hereafter denoted as \textit{R}) as shown in Fig.~\ref{fig:cav_highT}. This measurement configuration allows one to study the interference between reflected electrons which propagate towards Ohmic 3 directly without being affected by the cavity states, and those go through cavity and get modulated.    

In regime 1, where the reflector voltage V$_r$ was swept from - 0.30 (top trace) to - 0.27 V (bottom trace), a pronounced asymmetric resonance structure, which shows a peak at more negative arch-QPC voltage V$_a$ and valley at less negative end, centred around V$_a$= - 0.2 V was observed, which is also the centre of the 1D to 2D transition regime of arch-QPC  [as shown by a dotted rectangular box in Fig.~\ref{fig:cav_highT}(a)]. 

When the reflector was set to regime 2, V$_r$ increases from -0.27 to -0.24 V (Fig.~\ref{fig:cav_highT}(b)), the asymmetric resonance structure slowly evolves into a broad-shoulder structure, and a dip gradually forms at the pinch-off regime of the QPC. It is interesting to notice that the dip starts forming when the asymmetric resonance structure fully smears out. In addition, the centre of the shoulder structure shifts towards a more negative V$_a$ with increasing V$_r$ (see the solid line), on the contrary, the dip structure moves to a less negative V$_a$ (see the dashed line). 

In regime 3 (- 0.24 V $\leq$ V$_a$ $\leq$ 0), the intensity of both the shoulder structure and the dip decreases with less negative reflector voltage and finally smears out (Fig.~\ref{fig:cav_highT}(c) ).

  \begin{figure}
  	
  	\subfigure{    
  		\includegraphics[height=2.4in,width=3.6in]{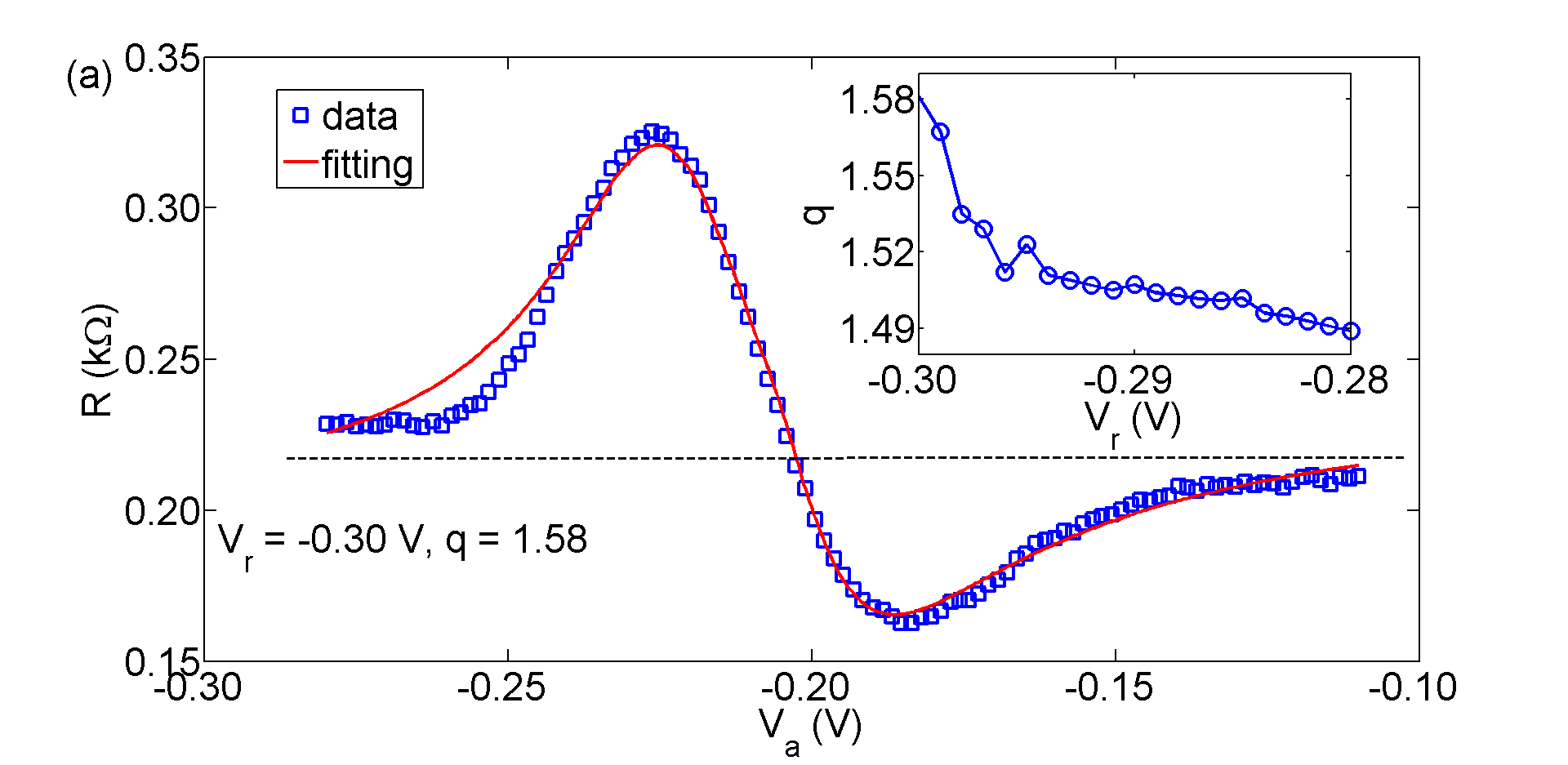} 
  		\label{fig:Fig3A } 
  	} % 
  	 \subfigure{
  		\includegraphics[height=2.4in,width=3.6in]{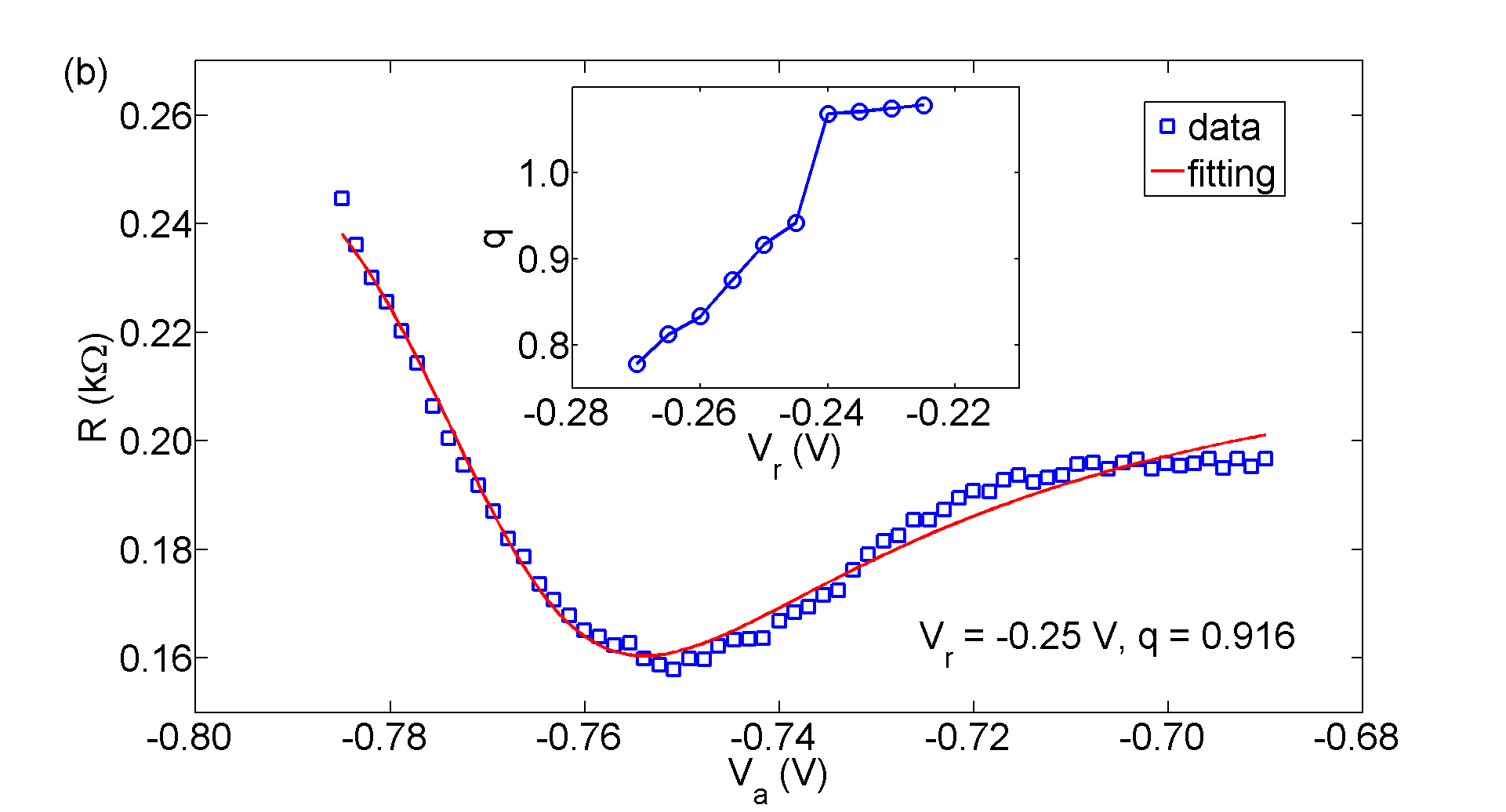}
  		\label{fig:Fig3B}
  	}% 

  	\caption[Fitting of the resonance structure and the dip ]{Theoretical fitting of the resonance structures (a) and dip (b). It is remarkable that both the resonance structure (V$_r$ = - 0.30 V) and the dip (V$_r$ = - 0.25 V) follow well defined Fano line shape. The horizontal black dotted line is a guide to the eye, reflecting the saturated value on the left end of the experimental data (the 1D regime of QPC) and that on the right (the 2D regime) do not align. This may be due to dramatic variation in DOS at the 1D-2D transition. Inset in plot (a) and (b) shows Fano factor q as a function of V$_r$ in the range where the structures are observable.  }
  	\label{fig:Fano_Va}
  \end{figure}
  
 The asymmetric resonance structure observed at the 1D-2D transition regime and the dip structure in the pinch-off regime seem to arise from dramatic change in the density of states (DOS) in the QPC in these regimes. However, such changes in DOS is independent of $V_r$ while the resonance and the dip structure are highly sensitive to $V_r$. In the present device, the angle of the reflector is properly designed to avoid back refection of the electrons, so that it is unlikely that reflected electrons would enter the QPC and perturb the DOS directly. 

A similar dip structure was reported in a double-QPC experiments when the intruder QPC was swept into pinch-off regime\cite{YKM09,FKY14}; the effect was attributed to Fano resonance which arises from the interference between the discrete states and the continuum\cite{FANO61}, where the role of the intruder was to provide the continuum. In our system, the electronic cavity which is defined using the arch-gate and the reflector, having its states filled up to chemical potential, hosts the continuum while the electrons injected from the arch-QPC are energetically discrete; thus the coupling between the QPC and the cavity states results in Fano resonance at the two regimes, (1) near the pinch-off and (2) at the 1D-2D transition. In addition, weak oscillations are observed in the resistance in the 1D regime of the arch-QPC (Fig.~\ref{fig:cav_highT}(a)) which could be a consequence of interference of 1D electrons with the cavity states. A detailed study in this regime will be published elsewhere\cite{CSM16}.

  \begin{figure}
  	
  	\subfigure{    
  		\includegraphics[height=2.4in,width=3.6in]{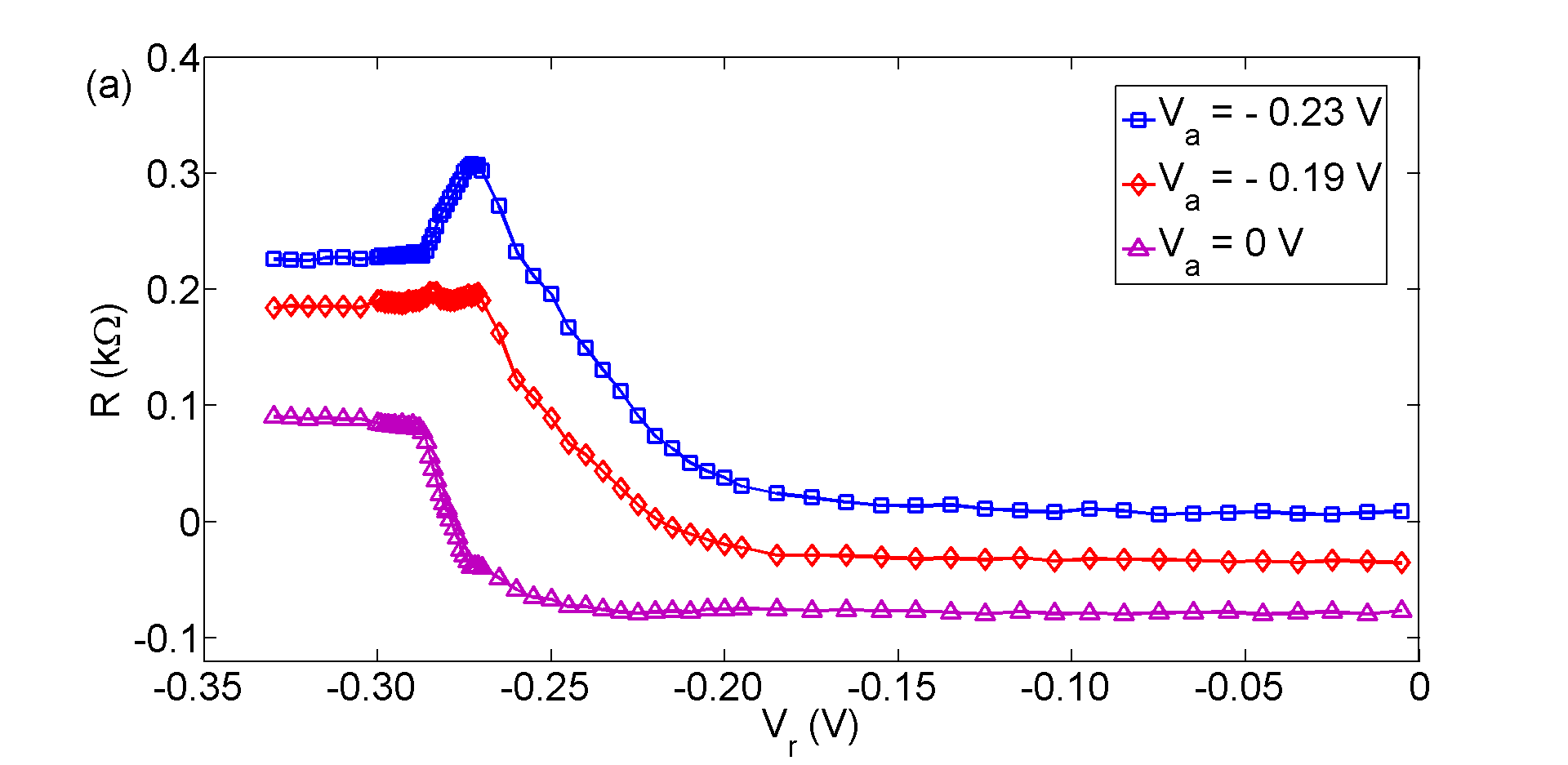} 
  		\label{fig:Fig4A} 
  	} % 
  	\subfigure{
  		\includegraphics[height=2.4in,width=3.6in]{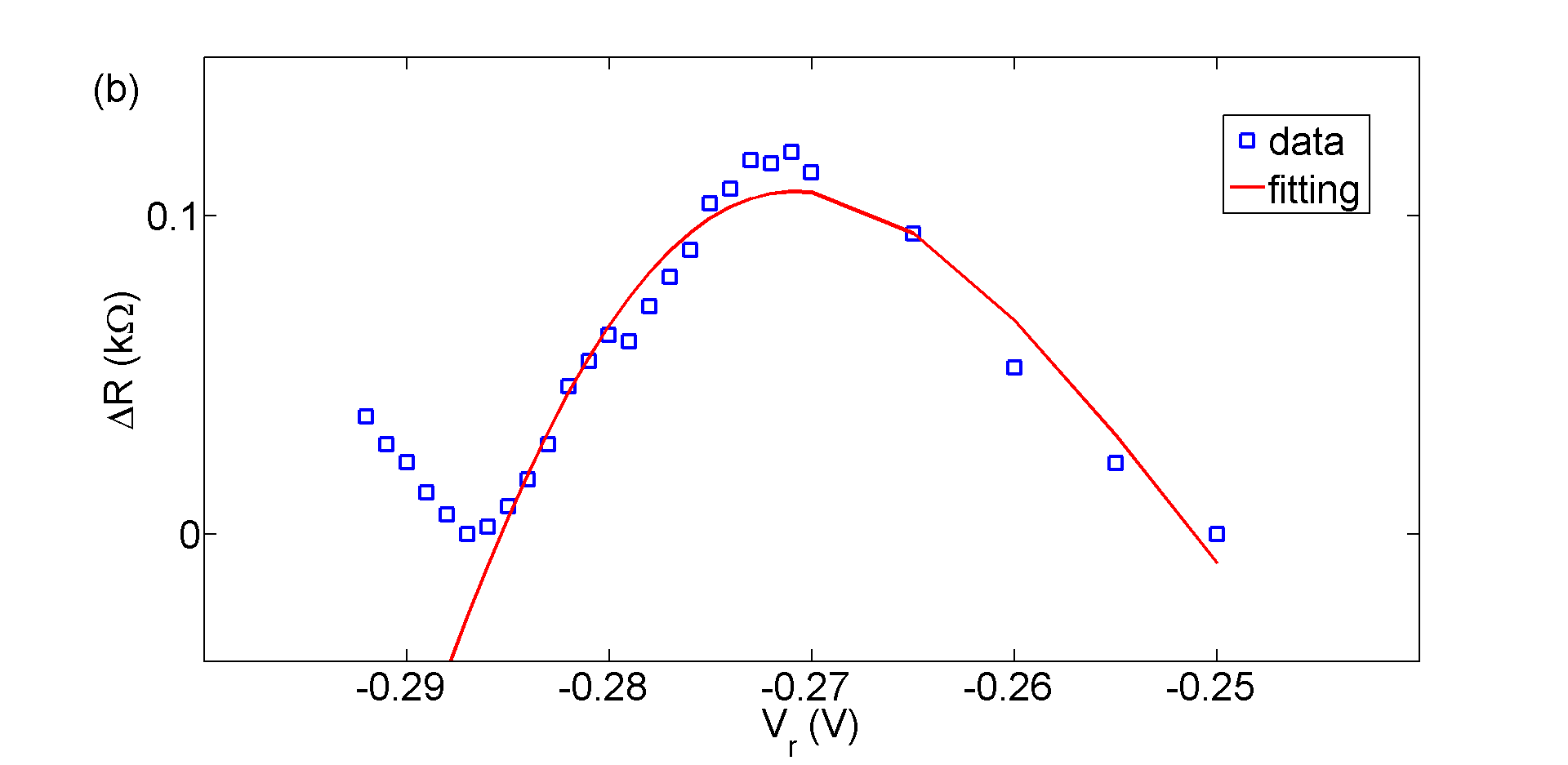}
  		\label{fig:Fig4B}
  	}% 

  	\caption[\textit{R} as a function of reflector voltage for various V$_a$]{\textit{R} as a function of reflector voltage for various V$_a$. (a) The arch-gate voltage was fixed at -0.23 V ($\square$), - 0.19 V ($\diamondsuit$) and 0 V ($\bigtriangleup$), respectively, while sweeping the reflector voltage; data have been offset vertically by 50 $\Omega$ for clarity. \textbf{b}, theoretical fitting of $\Delta$R(V$_r$,V$_a$) = R(V$_r$,V$_a$) - R(V$_r$,0), where V$_a$ = -0.23 V, using Eq.(1) where q = 1.47, it is clear that the data follow a Fano line shape.}
  	\label{fig:ref_fitting}
  \end{figure}
  
    \begin{figure}
    	
    	\subfigure{    
    		\includegraphics[height=2.4in,width=3.6in]{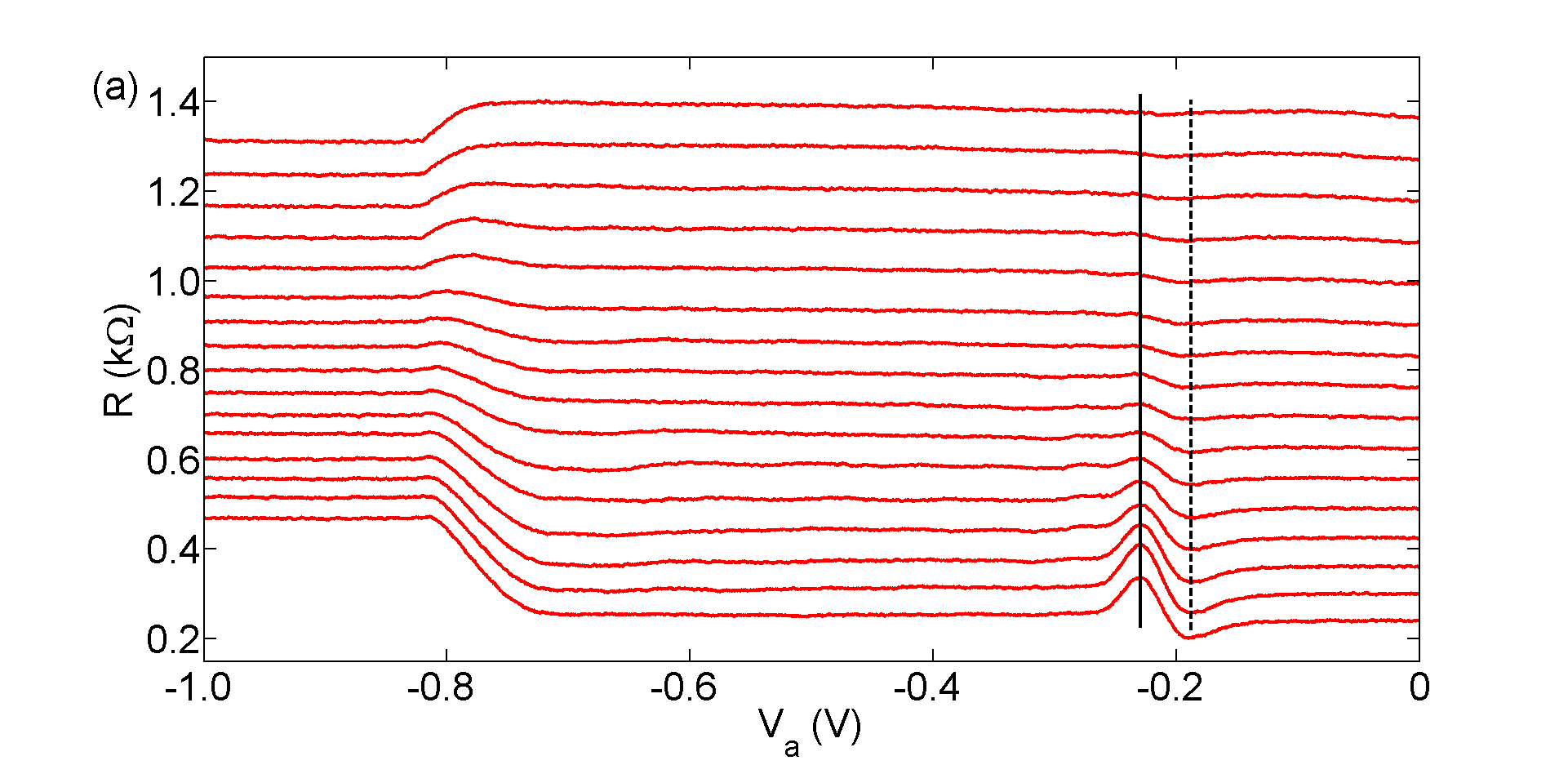} 
    		\label{fig:Fig5A} 
    	} % 
    	\subfigure{
    		\includegraphics[height=2.4in,width=3.6in]{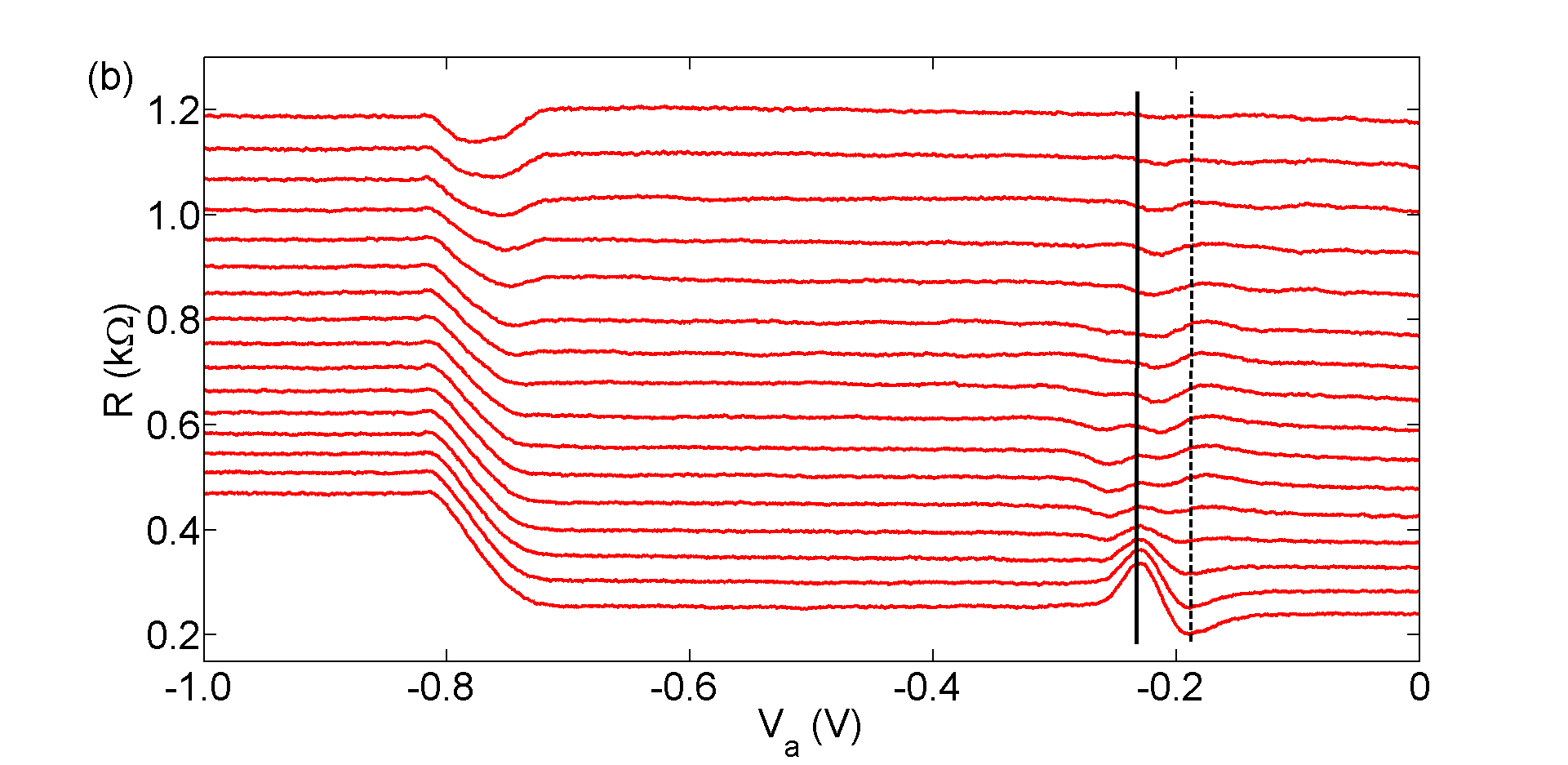}
    		\label{fig:Fig5B}
    	}% 

    	\caption[Effect of perpendicular magnetic field]{Effect of perpendicular magnetic field. In plot (a) and (b) the transverse magnetic field is incremented from 0 (bottom trace) to $\pm$ 150 mT (top trace), respectively, it is seen that both the peak (marked by the solid line) and the dip (highlighted by the dashed line) rapidly weaken against the field, they almost smears out at $\pm$ 100 mT. Data have been offset vertically by 20 $\Omega$ for clarity.  }
    	\label{fig:R1_Bdep}
    \end{figure} 
    
    \begin{figure}
    	
    	\subfigure{    
    		\includegraphics[height=2.4in,width=3.6in]{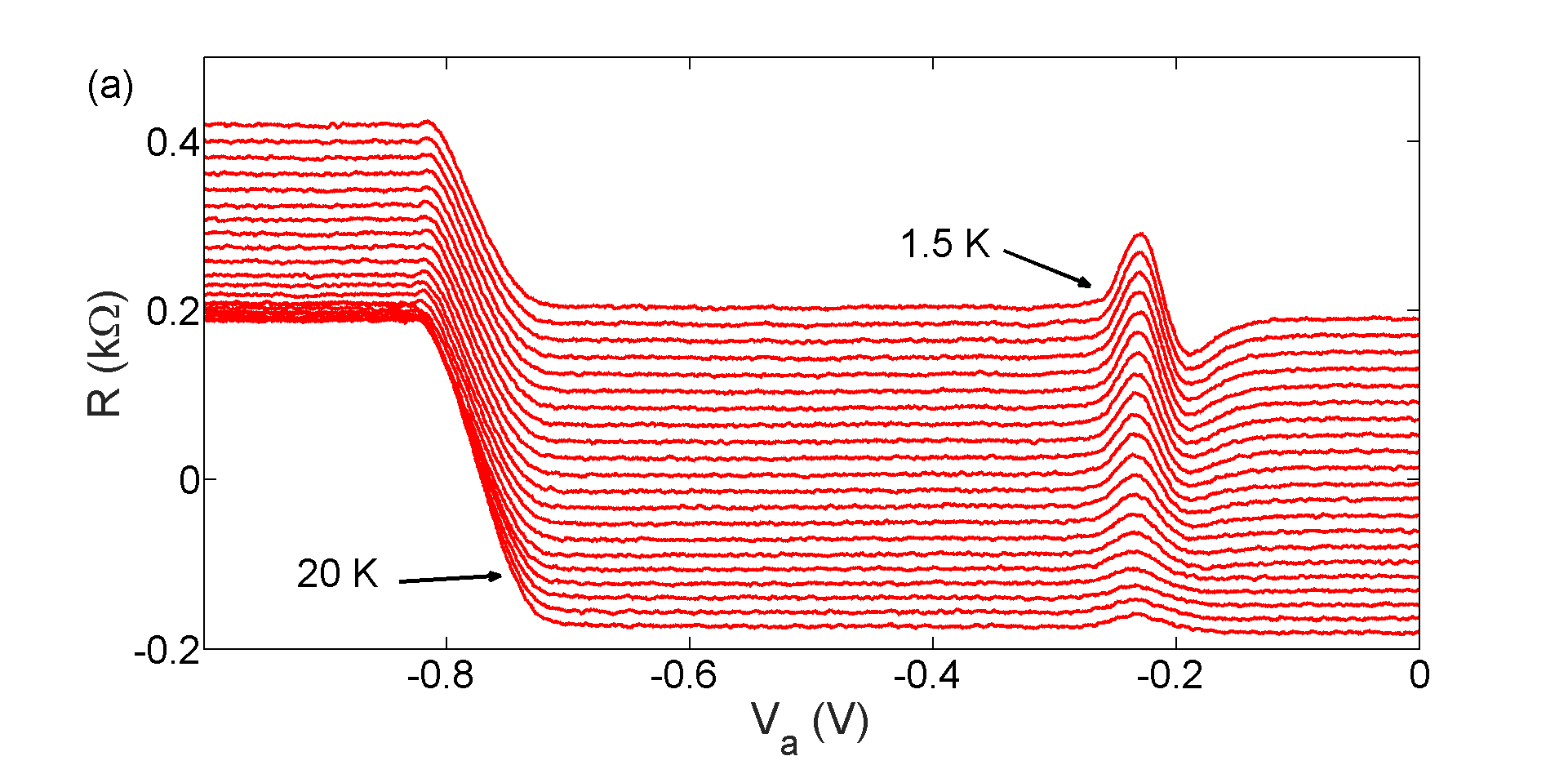} 
    		\label{fig:Fig6A} 
    	} % 
    	
    	\subfigure{
    		\includegraphics[height=2.4in,width=3.6in]{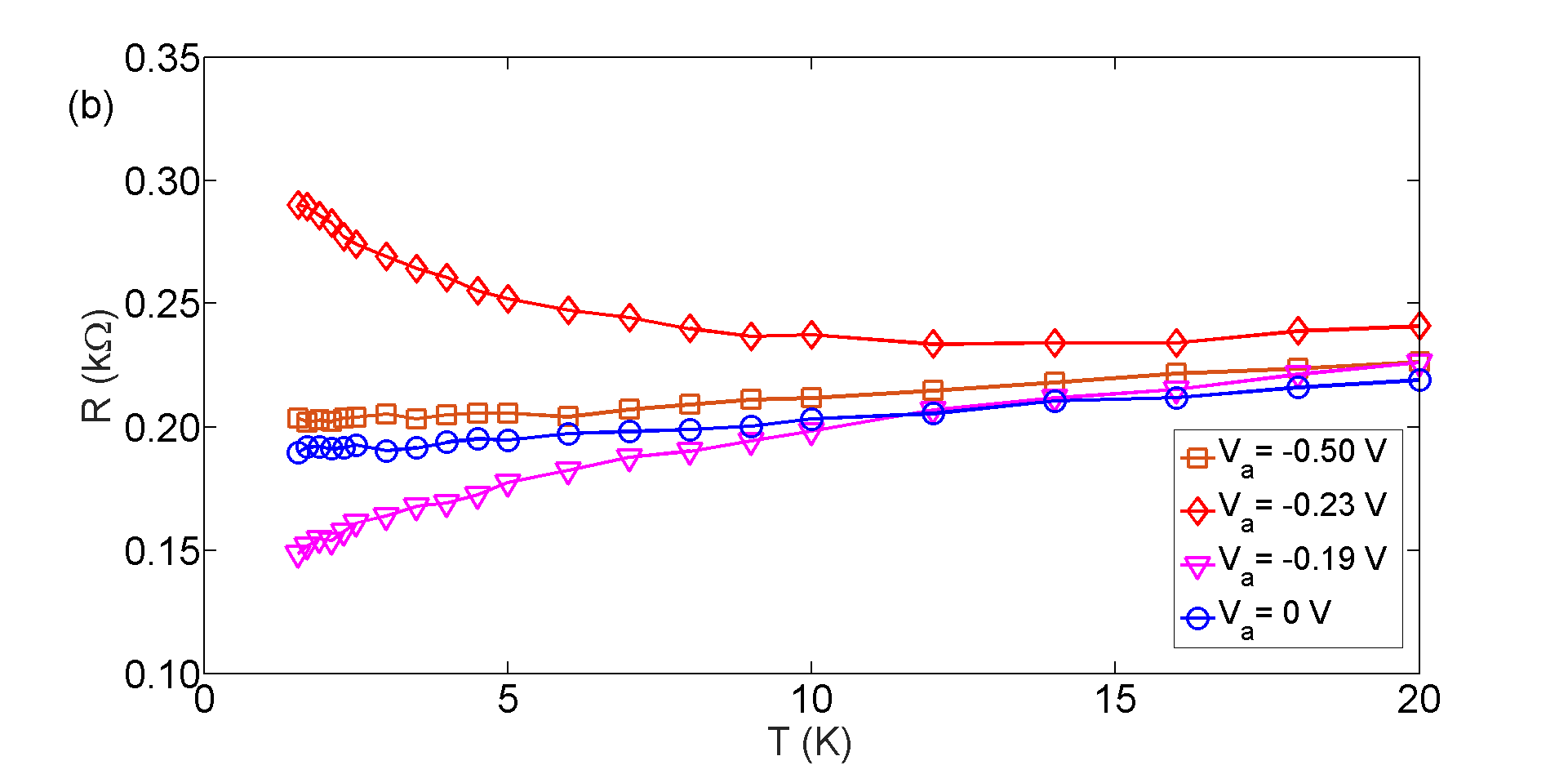}
    		\label{fig:Fig6B}
    	}% 
    	
    	\subfigure{
    		\includegraphics[height=2.4in,width=3.6in]{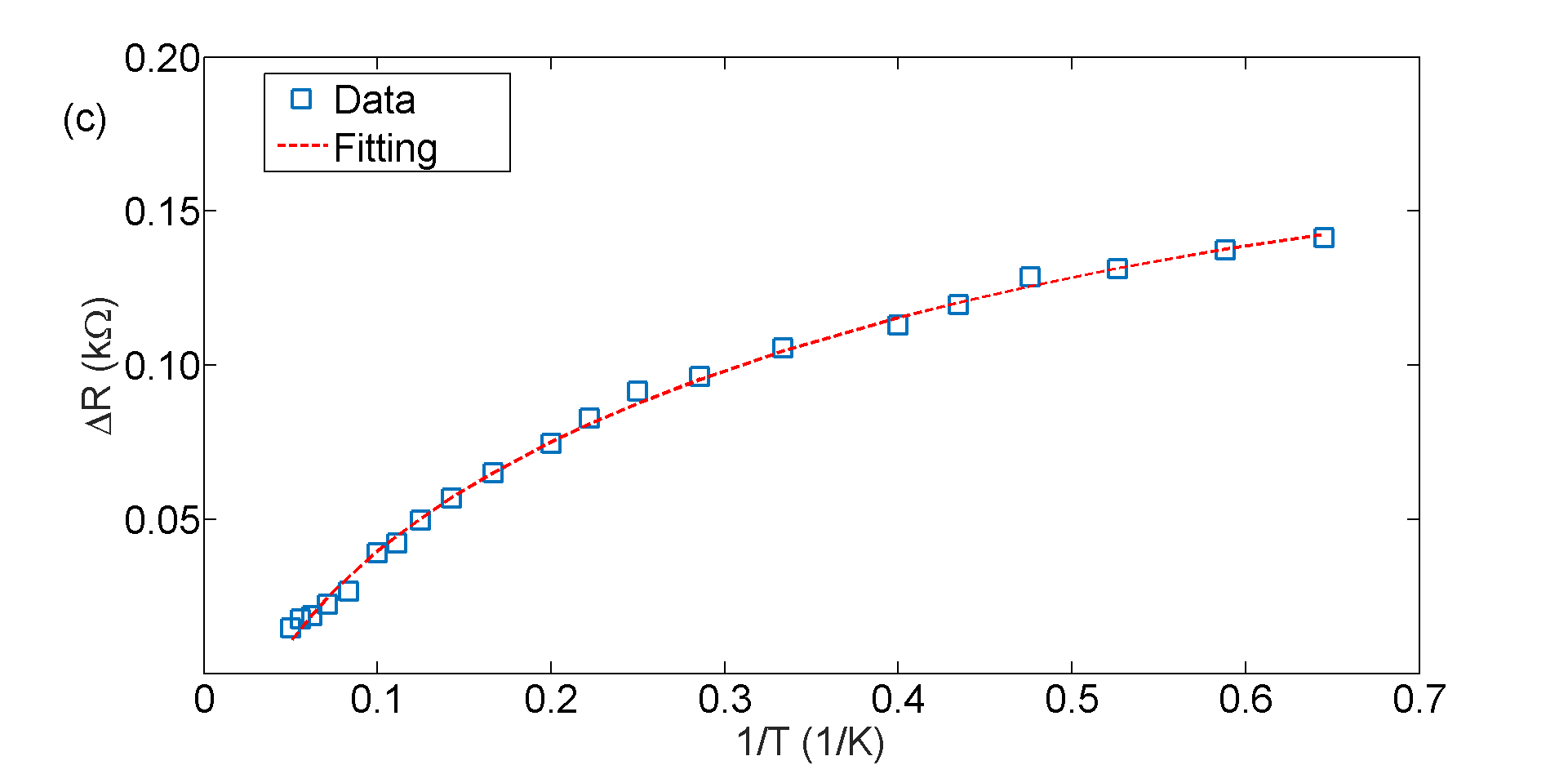}
    		\label{fig:Fig6C}
    	}% 

    	\caption[Temperature dependence of \textit{R}]{Temperature dependence of \textit{R}. (a)\textit{R} in regime 1 (V$_r$ = - 0.3 V) as a function of temperature from 1.5 K (top trace) to 20 K (bottom trace); it may be  seen that the intensity of both the peak and the dip structures decreases with increasing temperature. Data have been offset vertically by 20 $\Omega$ for clarity. (b) The temperature dependence of \textit{R} for V$_a$ = - 0.50  ($\square$), -  0.23  ($\diamondsuit$), - 0.19 ($\bigtriangledown$) and 0 V ($\bigcirc$); solid lines are a guide to the eye. (c) Mott fitting for temperature dependence of the peak structure, $\Delta R(V_a, T) = R(V_a, T) - R(0, T)$, where V$_a$ = - 0.23 V, R$_1$ = -0.2656 k$\Omega$, R$_2$ = 0.4238 k$\Omega$, T$_0$ = 1.2 K.  }
    	\label{fig:cav_T_dep}
    \end{figure} 

Figure~\ref{fig:Fano_Va} shows the fitting of the asymmetric resonance and the dip structures. It is remarkable that both the structures follow well defined Fano line shape\cite{FANO61,GHK00}, $$ R = R_0\frac{(q + \gamma (V - V_0))^2}{1 + \gamma^2(V - V_0)^2} + R_{inc}\eqno(1)$$ where R is the measured resistance, R$_0$ is a constant representing the amplitude of the resonance, q is the Fano factor which decides the asymmetry of the line shape, $\gamma = 20 V^{-1}$ (estimated from the Fermi energy and the pinch-off voltage), $V_0$ is the arch-gate voltage at the center of the resonance (dip), and R$_{inc}$ denotes the intrinsic contribution from the background\cite{GHK00}. However, it is noted that the fitting matches with the experimental result in the centre while it diverges at both ends for Fig. 3(a) and (b), which is likely due to the fact that \textit{R} changes as a function of V$_a$ even in the absence of the quantum correction, because electrons are collimated when the QPC is operating in the 1D regime, whereas they spread out when the QPC is in the 2D regime\cite{LAC90}. Moreover, the divergence in the pinch-off regime is due to the fact that once the 1D channel is pinched-off the detected signal will not change any further. The inset in Fig.~\ref{fig:Fano_Va}(a) and (b) shows Fano factor \textit{q} as a function of V$_r$. It was seen that \textit{q}, which represents the coupling between the QPC and the cavity, changes rapidly in both the regimes, which is most likely due to a sharp change in DOS of the QPC\cite{YKM09,FKY14} in the 1D-2D transition and pinch-off regime, respectively. Concerning the fitting in Fig.~3(b), the structure occurs in the pinch-off regime where making V$_a$ even more negative will not change the signal, therefore, this leaves an open question whether the data at the pinch-off are the dip of Fano resonance or do they a Breit-Wigner line shape. It is to be noted that an asymmetric Fano line shape is observed whenever resonant and nonresonant scattering paths interfere. However, when there is no interference between them, a symmetric Breit-Wigner resonance is expected\cite{GHK00}. If we look at Fig.~2(b) and focus on the dip in resistance at V$_a$ = - 0.8 V as a function of V$_r$, it seems that the dip structure takes a symmetric shape before finally settling at V$_r$=- 0.24 V to reflect an asymmetric line shape. It may be possible that the lower dip at the pinch-off is moving through the Breit-Wigner line shape to the Fano line shape, depending on the coupling factor q. However, such observations need a detailed study and it is difficult to comment on them at the moment.

The Fano line shape arising from coupling between the QPC and the cavity states can be modulated by either tuning the QPC states while fixing the cavity states as already shown in Fig.~\ref{fig:cav_highT}, or in a complementary way by adjusting the cavity states and holding the QPC states as shown in Fig.~\ref{fig:ref_fitting}(a). In this measurement, the reflector voltage was swept which allows one to control the energy spacing of the cavity states while $V_a$ was set to - 0.23 (the peak of asymmetric resonance structure), - 0.19 (the dip of asymmetric resonance structure) and 0 V, respectively, as shown in figure~\ref{fig:ref_fitting}(a). When $V_a$ was set to - 0.19 and 0 V,  respectively, where the cavity was not defined, it was seen that the resistance \textit{R} was initially almost constant when the reflector voltage $V_r > - 0.20$ V. On decreasing V$_r < -0.20$ V  a rise in resistance occurs simultaneously when the reflector conductance drops where the reflection probability \textit{r} increases rapidly, and then \textit{R} saturates when \textit{r} becomes unity (see inset, Fig.~\ref{fig:exp_setup}(b)). When V$_a$ was set to - 0.23 V , \textit{R} follows a similar trend as previous case when  $V_r > - 0.20$ V where the cavity was not activated. After the cavity was switched on, an anomalous peak in \textit{R} occurs when - 0.28 V $\leq$ V$_r$ $\leq$ - 0.25 V. The relative peak intensity, $\Delta$R(V$_r$,V$_a$) = R(V$_r$,V$_a$) - R(V$_r$,0), fits very well with Fano line shape, as shown in Fig.~\ref{fig:ref_fitting}(b) using Eq.(1), here R(V$_r$,0) is taken as background signal because it accounts for the change in \textit{R} due to the change in reflection probability only, and it is always present as the background signal regardless of V$_a$, and the resonance structure superposes on such a background; therefore, to highlight the resonance, we subtracted the background.

Coupling and decoupling between discrete and continuum states lead to Fano resonance, thus parameters which affect the coupling, apart from electrostatic confinement, are expected to influence the line shape dramatically\cite{FKY14}. Here, we present the effect of both negative and positive transverse magnetic field as shown in Fig.~\ref{fig:R1_Bdep}(a) and (b). The intensity of the asymmetric resonance structure is highly sensitive to the magnetic field in both field orientations, and almost smears out at a field of $\pm$ 100 mT. The reason for the the reduction of the intensity against magnetic field is three-fold: first, the chance of electron entering the cavity is magnetic field dependent; second, the cavity states, which represent the quantization of standing waves in the cavity, are highly dependent on the trajectory of electrons, with large magnetic field the electrons in the cavity become more localized and therefore coupling between 1D and cavity state is weakened; third, the perpendicular magnetic field also leads to a reduction of phase coherence length and thereby weakens the interference effect\cite{MMM81,MYH04}. A slight difference in the line shape for negative and positive field is likely to arise from the fact that both the inclined reflector and the negative field favor Ohmic 3, while a positive field prefers guiding electrons to Ohmic 4.  

We studied temperature dependence of \textit{R} as shown in Fig.~\ref{fig:cav_T_dep}. The reflector voltage was set to - 0.3 V, where both peak and dip of the asymmetric resonance structure are pronounced. It is found that the intensity of the peak and the dip decreases against increasing temperature, the dip structure smears out at around 16 K while the peak survives up to 20 K (although the height drops significantly). The evolution of line shape against temperature is clearer in Fig.~\ref{fig:cav_T_dep}(b), it is seen that \textit{R} increases slowly and almost linearly against temperature when V$_a$ is set to - 0.50 or 0 V, while the peak structure (V$_a$ = - 0.23 V) and the dip structure (V$_a$ = - 0.19 V ) change rapidly with increasing temperature.  The fitting of temperature dependence of relative intensity of the peak structure, $\Delta R(V_a, T) = R(V_a, T) - R(0, T)$ where V$_a$ = -0.23 V,  resembles a Mott line shape\cite{MOTT69} as shown in Fig.~\ref{fig:cav_T_dep}(c), using the relation, $$R=R_1 e^{(\frac{T_0}{T})^{\frac{1}{2}}}+R_2 e^{(\frac{T_0}{T})^{\frac{1}{3}}}\eqno(2)$$ where R$_1$, R$_2$ are fitting parameters while T$_0$  is defined as $$T_0 \propto \frac{1}{k_B N(E_F) \xi^3} \eqno(3)$$ where $k_B$ is Boltzmann constant, $N(E_F)$ is the density of state in the absence of electron-electron interaction and $\xi$ is the localization length\cite{MD79}, $T_0$ = 1.2 K is extracted from the fitting. The $\frac{1}{2}$ and $\frac{1}{3}$ terms arise from the contribution of 1D and 2D, respectively. This agrees with the fact the Fano resonance is a direct manifestation of coupling between the QPC and the cavity states.

Mott's law was initially proposed for random hopping in disordered system\cite{MOTT69}, whereas the clean systems generally do not follow Mott's law explicitly (e.g. see the trace for $V_a$ = 0 V in Fig.~\ref{fig:cav_T_dep}(b)). Random hopping accounts for the compromise between the spatial and energetic separation. It is possible that with the cavity switched on the multiple reflection process is a mimic of random hopping process, because the energy changing process in the cavity is accompanied with the spacial change\cite{HHH99} (states in the cavity are sensitive to the confinement length, the confinement length, in turn, is coordinate sensitive). 

In conclusions, we have shown the operation of a hybrid quantum device consisting of an arch-shaped QPC coupled to an electronic cavity, whose states can be tuned using a reflector gate. We have shown that, upon reducing the width of the QPC and increasing the reflection probability, Fano resonance can be produced at the 1D-2D transition and near the pinch off regime. The Fano resonant structure is very sensitive to the transverse magnetic field, and exhibits  Mott line-shape temperature dependence. Such systems show the potential of hybrid devices in realising tools for quantum information processing.

This work was supported by the Engineering and Physical Sciences Research Council (EPSRC), U.K.


\begin{thebibliography}{29}%
	\makeatletter
	\providecommand \@ifxundefined [1]{%
		\@ifx{#1\undefined}
	}%
	\providecommand \@ifnum [1]{%
		\ifnum #1\expandafter \@firstoftwo
		\else \expandafter \@secondoftwo
		\fi
	}%
	\providecommand \@ifx [1]{%
		\ifx #1\expandafter \@firstoftwo
		\else \expandafter \@secondoftwo
		\fi
	}%
	\providecommand \natexlab [1]{#1}%
	\providecommand \enquote  [1]{``#1''}%
	\providecommand \bibnamefont  [1]{#1}%
	\providecommand \bibfnamefont [1]{#1}%
	\providecommand \citenamefont [1]{#1}%
	\providecommand \href@noop [0]{\@secondoftwo}%
	\providecommand \href [0]{\begingroup \@sanitize@url \@href}%
	\providecommand \@href[1]{\@@startlink{#1}\@@href}%
	\providecommand \@@href[1]{\endgroup#1\@@endlink}%
	\providecommand \@sanitize@url [0]{\catcode `\\12\catcode `\$12\catcode
		`\&12\catcode `\#12\catcode `\^12\catcode `\_12\catcode `\%12\relax}%
	\providecommand \@@startlink[1]{}%
	\providecommand \@@endlink[0]{}%
	\providecommand \url  [0]{\begingroup\@sanitize@url \@url }%
	\providecommand \@url [1]{\endgroup\@href {#1}{\urlprefix }}%
	\providecommand \urlprefix  [0]{URL }%
	\providecommand \Eprint [0]{\href }%
	\providecommand \doibase [0]{http://dx.doi.org/}%
	\providecommand \selectlanguage [0]{\@gobble}%
	\providecommand \bibinfo  [0]{\@secondoftwo}%
	\providecommand \bibfield  [0]{\@secondoftwo}%
	\providecommand \translation [1]{[#1]}%
	\providecommand \BibitemOpen [0]{}%
	\providecommand \bibitemStop [0]{}%
	\providecommand \bibitemNoStop [0]{.\EOS\space}%
	\providecommand \EOS [0]{\spacefactor3000\relax}%
	\providecommand \BibitemShut  [1]{\csname bibitem#1\endcsname}%
	\let\auto@bib@innerbib\@empty
	%</preamble>
	\bibitem [{\citenamefont {Rauschenbeutel}\ \emph {et~al.}(1999)\citenamefont
		{Rauschenbeutel}, \citenamefont {Nogues}, \citenamefont {Osnaghi},
		\citenamefont {Bertet}, \citenamefont {Brune}, \citenamefont {Raimond},\ and\
		\citenamefont {Haroche}}]{RNO99}%
	\BibitemOpen
	\bibfield  {author} {\bibinfo {author} {\bibfnamefont {A.}~\bibnamefont
			{Rauschenbeutel}}, \bibinfo {author} {\bibfnamefont {G.}~\bibnamefont
			{Nogues}}, \bibinfo {author} {\bibfnamefont {S.}~\bibnamefont {Osnaghi}},
		\bibinfo {author} {\bibfnamefont {P.}~\bibnamefont {Bertet}}, \bibinfo
		{author} {\bibfnamefont {M.}~\bibnamefont {Brune}}, \bibinfo {author}
		{\bibfnamefont {J.~M.}\ \bibnamefont {Raimond}}, \ and\ \bibinfo {author}
		{\bibfnamefont {S.}~\bibnamefont {Haroche}},\ }\href {\doibase
		10.1103/PhysRevLett.83.5166} {\bibfield  {journal} {\bibinfo  {journal}
			{Phys. Rev. Lett.}\ }\textbf {\bibinfo {volume} {83}},\ \bibinfo {pages}
		{5166} (\bibinfo {year} {1999})}\BibitemShut {NoStop}%
	\bibitem [{\citenamefont {Hennessy}\ \emph {et~al.}(2007)\citenamefont
		{Hennessy}, \citenamefont {Badolato}, \citenamefont {Winger}, \citenamefont
		{Gerace}, \citenamefont {Atat{\"u}re}, \citenamefont {Gulde}, \citenamefont
		{F{\"a}lt}, \citenamefont {Hu},\ and\ \citenamefont
		{Imamo{\u{g}}lu}}]{HBW07}%
	\BibitemOpen
	\bibfield  {author} {\bibinfo {author} {\bibfnamefont {K.}~\bibnamefont
			{Hennessy}}, \bibinfo {author} {\bibfnamefont {A.}~\bibnamefont {Badolato}},
		\bibinfo {author} {\bibfnamefont {M.}~\bibnamefont {Winger}}, \bibinfo
		{author} {\bibfnamefont {D.}~\bibnamefont {Gerace}}, \bibinfo {author}
		{\bibfnamefont {M.}~\bibnamefont {Atat{\"u}re}}, \bibinfo {author}
		{\bibfnamefont {S.}~\bibnamefont {Gulde}}, \bibinfo {author} {\bibfnamefont
			{S.}~\bibnamefont {F{\"a}lt}}, \bibinfo {author} {\bibfnamefont {E.~L.}\
			\bibnamefont {Hu}}, \ and\ \bibinfo {author} {\bibfnamefont {A.}~\bibnamefont
			{Imamo{\u{g}}lu}},\ }\href@noop {} {\bibfield  {journal} {\bibinfo  {journal}
			{Nature}\ }\textbf {\bibinfo {volume} {445}},\ \bibinfo {pages} {896}
		(\bibinfo {year} {2007})}\BibitemShut {NoStop}%
	\bibitem [{\citenamefont {Niemczyk}\ \emph {et~al.}(2010)\citenamefont
		{Niemczyk}, \citenamefont {Deppe}, \citenamefont {Huebl}, \citenamefont
		{Menzel}, \citenamefont {Hocke}, \citenamefont {Schwarz}, \citenamefont
		{Garcia-Ripoll}, \citenamefont {Zueco}, \citenamefont {H{\"u}mmer},
		\citenamefont {Solano} \emph {et~al.}}]{NDH10}%
	\BibitemOpen
	\bibfield  {author} {\bibinfo {author} {\bibfnamefont {T.}~\bibnamefont
			{Niemczyk}}, \bibinfo {author} {\bibfnamefont {F.}~\bibnamefont {Deppe}},
		\bibinfo {author} {\bibfnamefont {H.}~\bibnamefont {Huebl}}, \bibinfo
		{author} {\bibfnamefont {E.}~\bibnamefont {Menzel}}, \bibinfo {author}
		{\bibfnamefont {F.}~\bibnamefont {Hocke}}, \bibinfo {author} {\bibfnamefont
			{M.}~\bibnamefont {Schwarz}}, \bibinfo {author} {\bibfnamefont
			{J.}~\bibnamefont {Garcia-Ripoll}}, \bibinfo {author} {\bibfnamefont
			{D.}~\bibnamefont {Zueco}}, \bibinfo {author} {\bibfnamefont
			{T.}~\bibnamefont {H{\"u}mmer}}, \bibinfo {author} {\bibfnamefont
			{E.}~\bibnamefont {Solano}},  \emph {et~al.},\ }\href@noop {} {\bibfield
		{journal} {\bibinfo  {journal} {Nature Physics}\ }\textbf {\bibinfo {volume}
			{6}},\ \bibinfo {pages} {772} (\bibinfo {year} {2010})}\BibitemShut {NoStop}%
	\bibitem [{\citenamefont {Duan}\ and\ \citenamefont {Kimble}(2004)}]{DK04}%
	\BibitemOpen
	\bibfield  {author} {\bibinfo {author} {\bibfnamefont {L.-M.}\ \bibnamefont
			{Duan}}\ and\ \bibinfo {author} {\bibfnamefont {H.~J.}\ \bibnamefont
			{Kimble}},\ }\href {\doibase 10.1103/PhysRevLett.92.127902} {\bibfield
		{journal} {\bibinfo  {journal} {Phys. Rev. Lett.}\ }\textbf {\bibinfo
			{volume} {92}},\ \bibinfo {pages} {127902} (\bibinfo {year}
		{2004})}\BibitemShut {NoStop}%
	\bibitem [{\citenamefont {Kwiat}\ \emph {et~al.}(1995)\citenamefont {Kwiat},
		\citenamefont {Mattle}, \citenamefont {Weinfurter}, \citenamefont
		{Zeilinger}, \citenamefont {Sergienko},\ and\ \citenamefont {Shih}}]{KMW95}%
	\BibitemOpen
	\bibfield  {author} {\bibinfo {author} {\bibfnamefont {P.~G.}\ \bibnamefont
			{Kwiat}}, \bibinfo {author} {\bibfnamefont {K.}~\bibnamefont {Mattle}},
		\bibinfo {author} {\bibfnamefont {H.}~\bibnamefont {Weinfurter}}, \bibinfo
		{author} {\bibfnamefont {A.}~\bibnamefont {Zeilinger}}, \bibinfo {author}
		{\bibfnamefont {A.~V.}\ \bibnamefont {Sergienko}}, \ and\ \bibinfo {author}
		{\bibfnamefont {Y.}~\bibnamefont {Shih}},\ }\href {\doibase
		10.1103/PhysRevLett.75.4337} {\bibfield  {journal} {\bibinfo  {journal}
			{Phys. Rev. Lett.}\ }\textbf {\bibinfo {volume} {75}},\ \bibinfo {pages}
		{4337} (\bibinfo {year} {1995})}\BibitemShut {NoStop}%
	\bibitem [{\citenamefont {Yoshie}\ \emph {et~al.}(2004)\citenamefont {Yoshie},
		\citenamefont {Scherer}, \citenamefont {Hendrickson}, \citenamefont
		{Khitrova}, \citenamefont {Gibbs}, \citenamefont {Rupper}, \citenamefont
		{Ell}, \citenamefont {Shchekin},\ and\ \citenamefont {Deppe}}]{YSJ04}%
	\BibitemOpen
	\bibfield  {author} {\bibinfo {author} {\bibfnamefont {T.}~\bibnamefont
			{Yoshie}}, \bibinfo {author} {\bibfnamefont {A.}~\bibnamefont {Scherer}},
		\bibinfo {author} {\bibfnamefont {J.}~\bibnamefont {Hendrickson}}, \bibinfo
		{author} {\bibfnamefont {G.}~\bibnamefont {Khitrova}}, \bibinfo {author}
		{\bibfnamefont {H.}~\bibnamefont {Gibbs}}, \bibinfo {author} {\bibfnamefont
			{G.}~\bibnamefont {Rupper}}, \bibinfo {author} {\bibfnamefont
			{C.}~\bibnamefont {Ell}}, \bibinfo {author} {\bibfnamefont {O.}~\bibnamefont
			{Shchekin}}, \ and\ \bibinfo {author} {\bibfnamefont {D.}~\bibnamefont
			{Deppe}},\ }\href@noop {} {\bibfield  {journal} {\bibinfo  {journal}
			{Nature}\ }\textbf {\bibinfo {volume} {432}},\ \bibinfo {pages} {200}
		(\bibinfo {year} {2004})}\BibitemShut {NoStop}%
	\bibitem [{\citenamefont {Carter}\ \emph {et~al.}(2013)\citenamefont {Carter},
		\citenamefont {Sweeney}, \citenamefont {Kim}, \citenamefont {Kim},
		\citenamefont {Solenov}, \citenamefont {Economou}, \citenamefont {Reinecke},
		\citenamefont {Yang}, \citenamefont {Bracker},\ and\ \citenamefont
		{Gammon}}]{CSK13}%
	\BibitemOpen
	\bibfield  {author} {\bibinfo {author} {\bibfnamefont {S.~G.}\ \bibnamefont
			{Carter}}, \bibinfo {author} {\bibfnamefont {T.~M.}\ \bibnamefont {Sweeney}},
		\bibinfo {author} {\bibfnamefont {M.}~\bibnamefont {Kim}}, \bibinfo {author}
		{\bibfnamefont {C.~S.}\ \bibnamefont {Kim}}, \bibinfo {author} {\bibfnamefont
			{D.}~\bibnamefont {Solenov}}, \bibinfo {author} {\bibfnamefont {S.~E.}\
			\bibnamefont {Economou}}, \bibinfo {author} {\bibfnamefont {T.~L.}\
			\bibnamefont {Reinecke}}, \bibinfo {author} {\bibfnamefont {L.}~\bibnamefont
			{Yang}}, \bibinfo {author} {\bibfnamefont {A.~S.}\ \bibnamefont {Bracker}}, \
		and\ \bibinfo {author} {\bibfnamefont {D.}~\bibnamefont {Gammon}},\
	}\href@noop {} {\bibfield  {journal} {\bibinfo  {journal} {Nature Photonics}\
	}\textbf {\bibinfo {volume} {7}},\ \bibinfo {pages} {329} (\bibinfo {year}
	{2013})}\BibitemShut {NoStop}%
\bibitem [{\citenamefont {Englund}\ \emph {et~al.}(2010)\citenamefont
	{Englund}, \citenamefont {Shields}, \citenamefont {Rivoire}, \citenamefont
	{Hatami}, \citenamefont {Vuckovic}, \citenamefont {Park},\ and\ \citenamefont
	{Lukin}}]{ESR10}%
\BibitemOpen
\bibfield  {author} {\bibinfo {author} {\bibfnamefont {D.}~\bibnamefont
		{Englund}}, \bibinfo {author} {\bibfnamefont {B.}~\bibnamefont {Shields}},
	\bibinfo {author} {\bibfnamefont {K.}~\bibnamefont {Rivoire}}, \bibinfo
	{author} {\bibfnamefont {F.}~\bibnamefont {Hatami}}, \bibinfo {author}
	{\bibfnamefont {J.}~\bibnamefont {Vuckovic}}, \bibinfo {author}
	{\bibfnamefont {H.}~\bibnamefont {Park}}, \ and\ \bibinfo {author}
	{\bibfnamefont {M.~D.}\ \bibnamefont {Lukin}},\ }\href@noop {} {\bibfield
	{journal} {\bibinfo  {journal} {Nano letters}\ }\textbf {\bibinfo {volume}
		{10}},\ \bibinfo {pages} {3922} (\bibinfo {year} {2010})}\BibitemShut
{NoStop}%
\bibitem [{\citenamefont {R\"ossler}\ \emph {et~al.}(2015)\citenamefont
	{R\"ossler}, \citenamefont {Oehri}, \citenamefont {Zilberberg}, \citenamefont
	{Blatter}, \citenamefont {Karalic}, \citenamefont {Pijnenburg}, \citenamefont
	{Hofmann}, \citenamefont {Ihn}, \citenamefont {Ensslin}, \citenamefont
	{Reichl},\ and\ \citenamefont {Wegscheider}}]{CDO15}%
\BibitemOpen
\bibfield  {author} {\bibinfo {author} {\bibfnamefont {C.}~\bibnamefont
		{R\"ossler}}, \bibinfo {author} {\bibfnamefont {D.}~\bibnamefont {Oehri}},
	\bibinfo {author} {\bibfnamefont {O.}~\bibnamefont {Zilberberg}}, \bibinfo
	{author} {\bibfnamefont {G.}~\bibnamefont {Blatter}}, \bibinfo {author}
	{\bibfnamefont {M.}~\bibnamefont {Karalic}}, \bibinfo {author} {\bibfnamefont
		{J.}~\bibnamefont {Pijnenburg}}, \bibinfo {author} {\bibfnamefont
		{A.}~\bibnamefont {Hofmann}}, \bibinfo {author} {\bibfnamefont
		{T.}~\bibnamefont {Ihn}}, \bibinfo {author} {\bibfnamefont {K.}~\bibnamefont
		{Ensslin}}, \bibinfo {author} {\bibfnamefont {C.}~\bibnamefont {Reichl}}, \
	and\ \bibinfo {author} {\bibfnamefont {W.}~\bibnamefont {Wegscheider}},\
}\href {\doibase 10.1103/PhysRevLett.115.166603} {\bibfield  {journal}
{\bibinfo  {journal} {Phys. Rev. Lett.}\ }\textbf {\bibinfo {volume} {115}},\
\bibinfo {pages} {166603} (\bibinfo {year} {2015})}\BibitemShut {NoStop}%
\bibitem [{\citenamefont {Wharam}\ \emph {et~al.}(1988)\citenamefont {Wharam},
	\citenamefont {Thornton}, \citenamefont {Newbury}, \citenamefont {Pepper},
	\citenamefont {Ahmed}, \citenamefont {Frost}, \citenamefont {Hasko},
	\citenamefont {Peacock}, \citenamefont {Ritchie},\ and\ \citenamefont
	{Jones}}]{DTM88}%
\BibitemOpen
\bibfield  {author} {\bibinfo {author} {\bibfnamefont {D.~A.}\ \bibnamefont
		{Wharam}}, \bibinfo {author} {\bibfnamefont {T.~J.}\ \bibnamefont
		{Thornton}}, \bibinfo {author} {\bibfnamefont {R.}~\bibnamefont {Newbury}},
	\bibinfo {author} {\bibfnamefont {M.}~\bibnamefont {Pepper}}, \bibinfo
	{author} {\bibfnamefont {H.}~\bibnamefont {Ahmed}}, \bibinfo {author}
	{\bibfnamefont {J.~E.~F.}\ \bibnamefont {Frost}}, \bibinfo {author}
	{\bibfnamefont {D.~G.}\ \bibnamefont {Hasko}}, \bibinfo {author}
	{\bibfnamefont {D.~C.}\ \bibnamefont {Peacock}}, \bibinfo {author}
	{\bibfnamefont {D.~A.}\ \bibnamefont {Ritchie}}, \ and\ \bibinfo {author}
	{\bibfnamefont {G.~A.~C.}\ \bibnamefont {Jones}},\ }\href@noop {} {\bibfield
	{journal} {\bibinfo  {journal} {Journal of Physics C: Solid State Physics}\
	}\textbf {\bibinfo {volume} {21}},\ \bibinfo {pages} {L209} (\bibinfo {year}
	{1988})}\BibitemShut {NoStop}%
\bibitem [{\citenamefont {van Wees}\ \emph {et~al.}(1988)\citenamefont {van
		Wees}, \citenamefont {van Houten}, \citenamefont {Beenakker}, \citenamefont
	{Williamson}, \citenamefont {Kouwenhoven}, \citenamefont {van~der Marel},\
	and\ \citenamefont {Foxon}}]{WHB88}%
\BibitemOpen
\bibfield  {author} {\bibinfo {author} {\bibfnamefont {B.~J.}\ \bibnamefont
		{van Wees}}, \bibinfo {author} {\bibfnamefont {H.}~\bibnamefont {van
			Houten}}, \bibinfo {author} {\bibfnamefont {C.~W.~J.}\ \bibnamefont
		{Beenakker}}, \bibinfo {author} {\bibfnamefont {J.~G.}\ \bibnamefont
		{Williamson}}, \bibinfo {author} {\bibfnamefont {L.~P.}\ \bibnamefont
		{Kouwenhoven}}, \bibinfo {author} {\bibfnamefont {D.}~\bibnamefont {van~der
			Marel}}, \ and\ \bibinfo {author} {\bibfnamefont {C.~T.}\ \bibnamefont
		{Foxon}},\ }\href {\doibase 10.1103/PhysRevLett.60.848} {\bibfield  {journal}
	{\bibinfo  {journal} {Phys. Rev. Lett.}\ }\textbf {\bibinfo {volume} {60}},\
	\bibinfo {pages} {848} (\bibinfo {year} {1988})}\BibitemShut {NoStop}%
\bibitem [{\citenamefont {Saito}\ \emph {et~al.}(1994)\citenamefont {Saito},
	\citenamefont {Usuki}, \citenamefont {Okada}, \citenamefont {Futatsugi},
	\citenamefont {Kiehl},\ and\ \citenamefont {Yokoyama}}]{MTM94}%
\BibitemOpen
\bibfield  {author} {\bibinfo {author} {\bibfnamefont {M.}~\bibnamefont
		{Saito}}, \bibinfo {author} {\bibfnamefont {T.}~\bibnamefont {Usuki}},
	\bibinfo {author} {\bibfnamefont {M.}~\bibnamefont {Okada}}, \bibinfo
	{author} {\bibfnamefont {T.}~\bibnamefont {Futatsugi}}, \bibinfo {author}
	{\bibfnamefont {R.~A.}\ \bibnamefont {Kiehl}}, \ and\ \bibinfo {author}
	{\bibfnamefont {N.}~\bibnamefont {Yokoyama}},\ }\href@noop {} {\bibfield
	{journal} {\bibinfo  {journal} {Applied Physics Letters}\ }\textbf {\bibinfo
		{volume} {65}} (\bibinfo {year} {1994})}\BibitemShut {NoStop}%
\bibitem [{\citenamefont {Katine}\ \emph {et~al.}(1997)\citenamefont {Katine},
	\citenamefont {Eriksson}, \citenamefont {Adourian}, \citenamefont
	{Westervelt}, \citenamefont {Edwards}, \citenamefont {Lupu-Sax},
	\citenamefont {Heller}, \citenamefont {Campman},\ and\ \citenamefont
	{Gossard}}]{KMW97}%
\BibitemOpen
\bibfield  {author} {\bibinfo {author} {\bibfnamefont {J.~A.}\ \bibnamefont
		{Katine}}, \bibinfo {author} {\bibfnamefont {M.~A.}\ \bibnamefont
		{Eriksson}}, \bibinfo {author} {\bibfnamefont {A.~S.}\ \bibnamefont
		{Adourian}}, \bibinfo {author} {\bibfnamefont {R.~M.}\ \bibnamefont
		{Westervelt}}, \bibinfo {author} {\bibfnamefont {J.~D.}\ \bibnamefont
		{Edwards}}, \bibinfo {author} {\bibfnamefont {A.}~\bibnamefont {Lupu-Sax}},
	\bibinfo {author} {\bibfnamefont {E.~J.}\ \bibnamefont {Heller}}, \bibinfo
	{author} {\bibfnamefont {K.~L.}\ \bibnamefont {Campman}}, \ and\ \bibinfo
	{author} {\bibfnamefont {A.~C.}\ \bibnamefont {Gossard}},\ }\href {\doibase
	10.1103/PhysRevLett.79.4806} {\bibfield  {journal} {\bibinfo  {journal}
		{Phys. Rev. Lett.}\ }\textbf {\bibinfo {volume} {79}},\ \bibinfo {pages}
	{4806} (\bibinfo {year} {1997})}\BibitemShut {NoStop}%
\bibitem [{\citenamefont {Duncan}\ \emph {et~al.}(2001)\citenamefont {Duncan},
	\citenamefont {Topinka}, \citenamefont {Westervelt}, \citenamefont
	{Maranowski},\ and\ \citenamefont {Gossard}}]{DTW01}%
\BibitemOpen
\bibfield  {author} {\bibinfo {author} {\bibfnamefont {D.~S.}\ \bibnamefont
		{Duncan}}, \bibinfo {author} {\bibfnamefont {M.~A.}\ \bibnamefont {Topinka}},
	\bibinfo {author} {\bibfnamefont {R.~M.}\ \bibnamefont {Westervelt}},
	\bibinfo {author} {\bibfnamefont {K.~D.}\ \bibnamefont {Maranowski}}, \ and\
	\bibinfo {author} {\bibfnamefont {A.~C.}\ \bibnamefont {Gossard}},\ }\href
{\doibase 10.1103/PhysRevB.64.033310} {\bibfield  {journal} {\bibinfo
		{journal} {Phys. Rev. B}\ }\textbf {\bibinfo {volume} {64}},\ \bibinfo
	{pages} {033310} (\bibinfo {year} {2001})}\BibitemShut {NoStop}%
\bibitem [{\citenamefont {D{\'\i}az}\ \emph {et~al.}(2005)\citenamefont
	{D{\'\i}az}, \citenamefont {Flores},\ and\ \citenamefont {Ponce}}]{DFP05}%
\BibitemOpen
\bibfield  {author} {\bibinfo {author} {\bibfnamefont {C.~A.~U.}\
		\bibnamefont {D{\'\i}az}}, \bibinfo {author} {\bibfnamefont {J.}~\bibnamefont
		{Flores}}, \ and\ \bibinfo {author} {\bibfnamefont {A.~P.}\ \bibnamefont
		{Ponce}},\ }\href@noop {} {\bibfield  {journal} {\bibinfo  {journal} {Solid
			state communications}\ }\textbf {\bibinfo {volume} {133}},\ \bibinfo {pages}
	{93} (\bibinfo {year} {2005})}\BibitemShut {NoStop}%
\bibitem [{YSM({\natexlab{a}})}]{YSM161}%
\BibitemOpen
\href@noop {} {\bibfield  {journal} {\bibinfo  {journal} {In Ref. [9], the
			dot-cavity hybrid system is dominated by the Kondo effect. However, in a
			cavity-reflctor hybrid quantum device, presented here, consisting of a QPC
			and an electronic cavity the Kondo effect is found to be absent, however, the
			device setup allows us to study the two-path interference effect which
			results in Fano resonance, a direct evidence of coupling between the 1D and
			cavity states}\ } }\BibitemShut {NoStop}%
\bibitem [{YSM({\natexlab{b}})}]{YSM162}%
\BibitemOpen
\href@noop {} {\bibfield  {journal} {\bibinfo  {journal} {The present device
			provides control over the relevant parameters required to study the coupling
			effect, however, due to limited number of gates, changes in the QPC states
			may be correlated with the cavity states. This limitation can be improved in
			future devices by pattering a local top gate over the QPC to adjust the QPC
			states without affecting the cavity states}\ } }\BibitemShut
{NoStop}%
\bibitem [{\citenamefont {Hersch}\ \emph {et~al.}(1999)\citenamefont {Hersch},
	\citenamefont {Haggerty},\ and\ \citenamefont {Heller}}]{HHH99}%
\BibitemOpen
\bibfield  {author} {\bibinfo {author} {\bibfnamefont {J.~S.}\ \bibnamefont
		{Hersch}}, \bibinfo {author} {\bibfnamefont {M.~R.}\ \bibnamefont
		{Haggerty}}, \ and\ \bibinfo {author} {\bibfnamefont {E.~J.}\ \bibnamefont
		{Heller}},\ }\href {\doibase 10.1103/PhysRevLett.83.5342} {\bibfield
	{journal} {\bibinfo  {journal} {Phys. Rev. Lett.}\ }\textbf {\bibinfo
		{volume} {83}},\ \bibinfo {pages} {5342} (\bibinfo {year}
	{1999})}\BibitemShut {NoStop}%
\bibitem [{\citenamefont {Hersch}\ \emph {et~al.}(2000)\citenamefont {Hersch},
	\citenamefont {Haggerty},\ and\ \citenamefont {Heller}}]{HHH00}%
\BibitemOpen
\bibfield  {author} {\bibinfo {author} {\bibfnamefont {J.~S.}\ \bibnamefont
		{Hersch}}, \bibinfo {author} {\bibfnamefont {M.~R.}\ \bibnamefont
		{Haggerty}}, \ and\ \bibinfo {author} {\bibfnamefont {E.~J.}\ \bibnamefont
		{Heller}},\ }\href {\doibase 10.1103/PhysRevE.62.4873} {\bibfield  {journal}
	{\bibinfo  {journal} {Phys. Rev. E}\ }\textbf {\bibinfo {volume} {62}},\
	\bibinfo {pages} {4873} (\bibinfo {year} {2000})}\BibitemShut {NoStop}%
\bibitem [{\citenamefont {Yoon}\ \emph {et~al.}(2009)\citenamefont {Yoon},
	\citenamefont {Kang}, \citenamefont {Morimoto}, \citenamefont {Mourokh},
	\citenamefont {Aoki}, \citenamefont {Reno}, \citenamefont {Bird},\ and\
	\citenamefont {Ochiai}}]{YKM09}%
\BibitemOpen
\bibfield  {author} {\bibinfo {author} {\bibfnamefont {Y.}~\bibnamefont
		{Yoon}}, \bibinfo {author} {\bibfnamefont {M.-G.}\ \bibnamefont {Kang}},
	\bibinfo {author} {\bibfnamefont {T.}~\bibnamefont {Morimoto}}, \bibinfo
	{author} {\bibfnamefont {L.}~\bibnamefont {Mourokh}}, \bibinfo {author}
	{\bibfnamefont {N.}~\bibnamefont {Aoki}}, \bibinfo {author} {\bibfnamefont
		{J.~L.}\ \bibnamefont {Reno}}, \bibinfo {author} {\bibfnamefont {J.~P.}\
		\bibnamefont {Bird}}, \ and\ \bibinfo {author} {\bibfnamefont
		{Y.}~\bibnamefont {Ochiai}},\ }\href {\doibase 10.1103/PhysRevB.79.121304}
{\bibfield  {journal} {\bibinfo  {journal} {Phys. Rev. B}\ }\textbf {\bibinfo
		{volume} {79}},\ \bibinfo {pages} {121304} (\bibinfo {year}
	{2009})}\BibitemShut {NoStop}%
\bibitem [{\citenamefont {Fransson}\ \emph {et~al.}(2014)\citenamefont
	{Fransson}, \citenamefont {Kang}, \citenamefont {Yoon}, \citenamefont {Xiao},
	\citenamefont {Ochiai}, \citenamefont {Reno}, \citenamefont {Aoki},\ and\
	\citenamefont {Bird}}]{FKY14}%
\BibitemOpen
\bibfield  {author} {\bibinfo {author} {\bibfnamefont {J.}~\bibnamefont
		{Fransson}}, \bibinfo {author} {\bibfnamefont {M.-G.}\ \bibnamefont {Kang}},
	\bibinfo {author} {\bibfnamefont {Y.}~\bibnamefont {Yoon}}, \bibinfo {author}
	{\bibfnamefont {S.}~\bibnamefont {Xiao}}, \bibinfo {author} {\bibfnamefont
		{Y.}~\bibnamefont {Ochiai}}, \bibinfo {author} {\bibfnamefont
		{J.}~\bibnamefont {Reno}}, \bibinfo {author} {\bibfnamefont {N.}~\bibnamefont
		{Aoki}}, \ and\ \bibinfo {author} {\bibfnamefont {J.}~\bibnamefont {Bird}},\
}\href@noop {} {\bibfield  {journal} {\bibinfo  {journal} {Nano letters}\
}\textbf {\bibinfo {volume} {14}},\ \bibinfo {pages} {788} (\bibinfo {year}
{2014})}\BibitemShut {NoStop}%
\bibitem [{\citenamefont {Fano}(1961)}]{FANO61}%
\BibitemOpen
\bibfield  {author} {\bibinfo {author} {\bibfnamefont {U.}~\bibnamefont
		{Fano}},\ }\href {\doibase 10.1103/PhysRev.124.1866} {\bibfield  {journal}
	{\bibinfo  {journal} {Phys. Rev.}\ }\textbf {\bibinfo {volume} {124}},\
	\bibinfo {pages} {1866} (\bibinfo {year} {1961})}\BibitemShut {NoStop}%
\bibitem [{\citenamefont {Yan}\ \emph {et~al.}(shed)\citenamefont {Yan},
	\citenamefont {Kumar}, \citenamefont {Pepper}, \citenamefont {See},
	\citenamefont {Farrer}, \citenamefont {Ritchie}, \citenamefont {Griffiths},\
	and\ \citenamefont {Jones}}]{CSM16}%
\BibitemOpen
\bibfield  {author} {\bibinfo {author} {\bibfnamefont {C.}~\bibnamefont
		{Yan}}, \bibinfo {author} {\bibfnamefont {S.}~\bibnamefont {Kumar}}, \bibinfo
	{author} {\bibfnamefont {M.}~\bibnamefont {Pepper}}, \bibinfo {author}
	{\bibfnamefont {P.}~\bibnamefont {See}}, \bibinfo {author} {\bibfnamefont
		{I.}~\bibnamefont {Farrer}}, \bibinfo {author} {\bibfnamefont
		{D.}~\bibnamefont {Ritchie}}, \bibinfo {author} {\bibfnamefont
		{J.}~\bibnamefont {Griffiths}}, \ and\ \bibinfo {author} {\bibfnamefont
		{G.}~\bibnamefont {Jones}},\ }\href@noop {} {\  (\bibinfo {year} {to be
		published})}\BibitemShut {NoStop}%
\bibitem [{\citenamefont {G\"ores}\ \emph {et~al.}(2000)\citenamefont
	{G\"ores}, \citenamefont {Goldhaber-Gordon}, \citenamefont {Heemeyer},
	\citenamefont {Kastner}, \citenamefont {Shtrikman}, \citenamefont {Mahalu},\
	and\ \citenamefont {Meirav}}]{GHK00}%
\BibitemOpen
\bibfield  {author} {\bibinfo {author} {\bibfnamefont {J.}~\bibnamefont
		{G\"ores}}, \bibinfo {author} {\bibfnamefont {D.}~\bibnamefont
		{Goldhaber-Gordon}}, \bibinfo {author} {\bibfnamefont {S.}~\bibnamefont
		{Heemeyer}}, \bibinfo {author} {\bibfnamefont {M.~A.}\ \bibnamefont
		{Kastner}}, \bibinfo {author} {\bibfnamefont {H.}~\bibnamefont {Shtrikman}},
	\bibinfo {author} {\bibfnamefont {D.}~\bibnamefont {Mahalu}}, \ and\ \bibinfo
	{author} {\bibfnamefont {U.}~\bibnamefont {Meirav}},\ }\href {\doibase
	10.1103/PhysRevB.62.2188} {\bibfield  {journal} {\bibinfo  {journal} {Phys.
			Rev. B}\ }\textbf {\bibinfo {volume} {62}},\ \bibinfo {pages} {2188}
	(\bibinfo {year} {2000})}\BibitemShut {NoStop}%
\bibitem [{\citenamefont {Molenkamp}\ \emph {et~al.}(1990)\citenamefont
	{Molenkamp}, \citenamefont {Staring}, \citenamefont {Beenakker},
	\citenamefont {Eppenga}, \citenamefont {Timmering}, \citenamefont
	{Williamson}, \citenamefont {Harmans},\ and\ \citenamefont {Foxon}}]{LAC90}%
\BibitemOpen
\bibfield  {author} {\bibinfo {author} {\bibfnamefont {L.~W.}\ \bibnamefont
		{Molenkamp}}, \bibinfo {author} {\bibfnamefont {A.~A.~M.}\ \bibnamefont
		{Staring}}, \bibinfo {author} {\bibfnamefont {C.~W.~J.}\ \bibnamefont
		{Beenakker}}, \bibinfo {author} {\bibfnamefont {R.}~\bibnamefont {Eppenga}},
	\bibinfo {author} {\bibfnamefont {C.~E.}\ \bibnamefont {Timmering}}, \bibinfo
	{author} {\bibfnamefont {J.~G.}\ \bibnamefont {Williamson}}, \bibinfo
	{author} {\bibfnamefont {C.~J. P.~M.}\ \bibnamefont {Harmans}}, \ and\
	\bibinfo {author} {\bibfnamefont {C.~T.}\ \bibnamefont {Foxon}},\ }\href
{\doibase 10.1103/PhysRevB.41.1274} {\bibfield  {journal} {\bibinfo
		{journal} {Phys. Rev. B}\ }\textbf {\bibinfo {volume} {41}},\ \bibinfo
	{pages} {1274} (\bibinfo {year} {1990})}\BibitemShut {NoStop}%
\bibitem [{\citenamefont {Kaveh}\ \emph {et~al.}(1981)\citenamefont {Kaveh},
	\citenamefont {Uren}, \citenamefont {Davies},\ and\ \citenamefont
	{Pepper}}]{MMM81}%
\BibitemOpen
\bibfield  {author} {\bibinfo {author} {\bibfnamefont {M.}~\bibnamefont
		{Kaveh}}, \bibinfo {author} {\bibfnamefont {M.~J.}\ \bibnamefont {Uren}},
	\bibinfo {author} {\bibfnamefont {R.~A.}\ \bibnamefont {Davies}}, \ and\
	\bibinfo {author} {\bibfnamefont {M.}~\bibnamefont {Pepper}},\ }\href@noop {}
{\bibfield  {journal} {\bibinfo  {journal} {Journal of Physics C: Solid State
			Physics}\ }\textbf {\bibinfo {volume} {14}},\ \bibinfo {pages} {L413}
	(\bibinfo {year} {1981})}\BibitemShut {NoStop}%
\bibitem [{\citenamefont {McPhail}\ \emph {et~al.}(2004)\citenamefont
	{McPhail}, \citenamefont {Yasin}, \citenamefont {Hamilton}, \citenamefont
	{Simmons}, \citenamefont {Linfield}, \citenamefont {Pepper},\ and\
	\citenamefont {Ritchie}}]{MYH04}%
\BibitemOpen
\bibfield  {author} {\bibinfo {author} {\bibfnamefont {S.}~\bibnamefont
		{McPhail}}, \bibinfo {author} {\bibfnamefont {C.~E.}\ \bibnamefont {Yasin}},
	\bibinfo {author} {\bibfnamefont {A.~R.}\ \bibnamefont {Hamilton}}, \bibinfo
	{author} {\bibfnamefont {M.~Y.}\ \bibnamefont {Simmons}}, \bibinfo {author}
	{\bibfnamefont {E.~H.}\ \bibnamefont {Linfield}}, \bibinfo {author}
	{\bibfnamefont {M.}~\bibnamefont {Pepper}}, \ and\ \bibinfo {author}
	{\bibfnamefont {D.~A.}\ \bibnamefont {Ritchie}},\ }\href {\doibase
	10.1103/PhysRevB.70.245311} {\bibfield  {journal} {\bibinfo  {journal} {Phys.
			Rev. B}\ }\textbf {\bibinfo {volume} {70}},\ \bibinfo {pages} {245311}
	(\bibinfo {year} {2004})}\BibitemShut {NoStop}%
\bibitem [{\citenamefont {Mott}(1969)}]{MOTT69}%
\BibitemOpen
\bibfield  {author} {\bibinfo {author} {\bibfnamefont {N.}~\bibnamefont
		{Mott}},\ }\href@noop {} {\bibfield  {journal} {\bibinfo  {journal}
		{Philosophical Magazine}\ }\textbf {\bibinfo {volume} {19}},\ \bibinfo
	{pages} {835} (\bibinfo {year} {1969})}\BibitemShut {NoStop}%
\bibitem [{\citenamefont {Mott}\ and\ \citenamefont {Davis}(1979)}]{MD79}%
\BibitemOpen
\bibfield  {author} {\bibinfo {author} {\bibfnamefont {N.}~\bibnamefont
		{Mott}}\ and\ \bibinfo {author} {\bibfnamefont {E.}~\bibnamefont {Davis}},\
}\href@noop {} {\enquote {\bibinfo {title} {Electronic processes in
		non-crystalline materials, 2nd edn. clarendon},}\ } (\bibinfo {year}
{1979})\BibitemShut {NoStop}%
\end{thebibliography}
\end{document}